\documentclass[journal,a4paper]{IEEEtran}

\usepackage[]{graphicx}
\usepackage{url}
\usepackage{ifpdf}
\usepackage{hyperref}
\hypersetup{breaklinks,colorlinks,citecolor=blue}
\usepackage[noadjust]{cite}
\usepackage{tikz} 
\usepackage{pgfplots} 
\usetikzlibrary{arrows} 
\usepackage{cite}
\usepackage{amssymb}
\usepackage{amsmath}
\usepackage{mathtools}

\DeclarePairedDelimiter{\ceil}{\lceil}{\rceil}
\DeclarePairedDelimiter{\floor}{\lfloor}{\rfloor}

\usepackage{subfigure}
\usepackage{color}
\usepackage{pstool}

\newcommand{\eqn}[1]{(#1)}

\newcommand{\fig}[1]{Fig.~#1}

\newcommand{\sectn}[1]{Sec.~#1}

\newcommand{\eg}{\mbox{\it e.g.}}
\newcommand{\ie}{\mbox{\it i.e.}}

\newcommand{\cf}{\mbox{\it cf.}}

\newcommand{\healpix}{{\tt HEALPix}}

\newcommand{\sshtcode}{{\tt SSHT}}
\newcommand{\sothreecode}{{\tt SO3}}
\newcommand{\stwoletcode}{{\tt S2LET}}

\newcommand{\spcend}{\ensuremath{\:}}
\newcommand{\img}{\ensuremath{{\rm i}}}
\newcommand{\cconj}{\ensuremath{\ast}} 
 
\newcommand{\reals}{\ensuremath{\mathbb{R}}}

\newcommand{\realsnz}{\ensuremath{\mathbb{R}^{+}_{\ast}}}
\newcommand{\integers}{\ensuremath{\mathbb{Z}}}
\newcommand{\naturals}{\ensuremath{\mathbb{N}}}

\newcommand{\ltwo}{\ensuremath{\mathrm{L}^2}}
\newcommand{\sphere}{\ensuremath{{\mathbb{S}^2}}}
\newcommand{\sothree}{\ensuremath{{\mathrm{SO}(3)}}}

\newcommand{\dx}{\ensuremath{\mathrm{\,d}}}
\newcommand{\dmu}[1]{\ensuremath{\dx \Omega(#1)}}

\newcommand{\deul}[1]{\ensuremath{\dx \varrho(#1)}}

\newcommand{\innerp}[2]{\ensuremath{\langle {#1},\: {#2} \rangle}}

\newcommand{\sa}{\ensuremath{\omega}}
\newcommand{\saa}{\ensuremath{\theta}}
\newcommand{\sab}{\ensuremath{\varphi}}
\newcommand{\sas}{\ensuremath{\saa, \sab}}
\newcommand{\eul}{\ensuremath{\mathbf{\rho}}}
\newcommand{\euls}{\ensuremath{\eula, \eulb, \eulc}}
\newcommand{\eula}{\ensuremath{\alpha}}
\newcommand{\eulb}{\ensuremath{\beta}}
\newcommand{\eulc}{\ensuremath{\gamma}}
\newcommand{\eulai}{\ensuremath{a}}
\newcommand{\eulbi}{\ensuremath{b}}
\newcommand{\eulci}{\ensuremath{g}}
\newcommand{\eulaiang}{\ensuremath{\eula_\eulai}}
\newcommand{\eulbiang}{\ensuremath{\eulb_\eulbi}}
\newcommand{\eulciang}{\ensuremath{\eulc_\eulci}}
\newcommand{\el}{\ensuremath{\ell}}
\newcommand{\m}{\ensuremath{m}}
\newcommand{\n}{\ensuremath{n}}

\newcommand{\spin}{\ensuremath{s}}

\newcommand{\elmax}{\ensuremath{{L}}}

\newcommand{\nmax}{\ensuremath{{N}}}

\newcommand{\p}{\ensuremath{^\prime}}
\newcommand{\pp}{\ensuremath{^{\prime\prime}}}

\newcommand{\kron}[2]{\ensuremath{\delta_{{#1}{#2}}}}

\renewcommand{\exp}[1]{\ensuremath{{\rm e}^{#1}}}

\newcommand{\sshfarg}[4]{\ensuremath{{{}_{#4} Y_{#1#2}({#3})}}}
\newcommand{\sshfargc}[4]{\ensuremath{{{}_{#4} Y_{#1#2}^\cconj({#3})}}}
\newcommand{\sshfargsp}[4]{\ensuremath{{{}_{#4} Y_{{#1},{#2}}({#3})}}}
\newcommand{\shf}[2]{\ensuremath{Y_{#1#2}}}

\newcommand{\shc}[3]{\ensuremath{{#1}_{{#2}{#3}}}}

\newcommand{\sshf}[3]{\ensuremath{{{}_{#3} Y_{#1#2}}}}

\newcommand{\sshc}[4]{\ensuremath{{}_{#4} {#1}_{{#2}{#3}}}}

\newcommand{\dmatbig}{\ensuremath{D}}
\newcommand{\Dlmn}{\ensuremath{ \dmatbig_{\m\n}^{\el} }}
\newcommand{\Dlmnc}{\ensuremath{ \dmatbig_{\m\n}^{\el\cconj} }}

\newcommand{\Dlmnpc}{\ensuremath{ \dmatbig_{\m\n}^{\el\cconj}(\eul) }}

\newcommand{\dmatsmall}{\ensuremath{d}}
\newcommand{\dlmn}{\ensuremath{ \dmatsmall_{\m\n}^{\el} }}

\newcommand{\dlmnb}{\ensuremath{ \dmatsmall_{\m\n}^{\el}(\eulb) }}

\newcommand{\dlmnhalfpi}[3]{\ensuremath{ \Delta_{{#2}{#3}}^{#1} }}

\newcommand{\wigc}[4]{\ensuremath{{#1}^{#2}_{{#3}{#4}}}}

\newcommand{\rotarg}[1]{\ensuremath{\mathcal{R}_{#1}}}

\newcommand{\f}{\ensuremath{f}}
\newcommand{\fs}{\ensuremath{{}_\spin f}}

\newcommand{\fslm}{\ensuremath{\shc{\fs}{\el}{\m}}}

\newcommand{\wav}{\ensuremath{\psi}}
\newcommand{\swav}{\ensuremath{{}_\spin\psi}}

\newcommand{\wavs}{\ensuremath{\Phi}}
\newcommand{\swavs}{\ensuremath{{}_\spin\Phi}}
\newcommand{\wcoeff}{\ensuremath{W}}
\newcommand{\scoeff}{\ensuremath{W}}
\newcommand{\wscale}{\ensuremath{j}}
\newcommand{\wscalemax}{\ensuremath{J}}
\newcommand{\wscalemin}{\ensuremath{J_0}}

\newcommand{\dilparam}{\ensuremath{\alpha}}
\newcommand{\wavker}{\ensuremath{\kappa}}
\newcommand{\wavsteer}{\ensuremath{s}}

\newcommand{\sumlm}{\ensuremath{\sum_{\el=0}^{\infty} \sum_{\m=-\el}^\el}}
\newcommand{\sumlmn}{\ensuremath{\sum_{\el=0}^{\infty} \sum_{\m=-\el}^\el} \sum_{\n=-\el}^\el}

\newcommand{\summ}{\ensuremath{\sum_{\m=-\el}^\el}}

\newcommand{\sumn}{\ensuremath{\sum_{\n=-\el}^\el}}

\newcommand{\weight}{\ensuremath{w}}
\newcommand{\order}{\ensuremath{\mathcal{O}}}

\renewcommand{\eqn}[1]{Eqn.~(#1)}

\renewcommand{\wav}{\ensuremath{\Psi}}
\renewcommand{\swav}{\ensuremath{{}_\spin\Psi}}
\renewcommand{\wavsteer}{\ensuremath{\zeta}}
\renewcommand{\dilparam}{\ensuremath{{\lambda}}}

\renewcommand{\sshtcode}{{\sc ssht}}
\renewcommand{\sothreecode}{{\sc so3}}
\renewcommand{\stwoletcode}{{\sc s2let}}
\renewcommand{\healpix}{{\sc healp}{\small ix}}
\newcommand{\fftwcode}{{\sc fftw}}

\begin{document}
%
\title{Second-Generation Curvelets on the Sphere}
%
%
%
\author{Jennifer Y.~H.~Chan, Boris Leistedt, Thomas D.~Kitching, Jason D.~McEwen 
 \thanks{Copyright (c) 2015 IEEE. Personal use of this material is permitted. However, permission to use this material for any other purposes must be obtained from the IEEE by sending a request to pubs-permissions@ieee.org.}
  \thanks{J.~Y.~H.~Chan was supported by the 
   UCL Graduate Student Scholarship and Overseas Student Scholarship. 
    B.~Leistedt was supported by the
    European Research Council under the European Community’s Seventh
    Framework Programme (FP7/2007-2013) / ERC grant agreement no
    306478-CosmicDawn, and by the Impact and Perren funds. 
    T.~D.~Kitching was supported by a Royal Society URF. 
    J.~D.~McEwen was supported by the Engineering and Physical
    Sciences Research Council (grant number EP/M011852/1).}
  \thanks{J.~Y.~H.~Chan, T.~D.~Kitching and J.~D.~McEwen are with the Mullard Space Science Laboratory
    (MSSL), University College London (UCL), Surrey RH5 6NT, UK.
    B.~Leistedt is with the Department 
    of Physics and Astronomy, UCL, London WC1E 6BT, UK, 
    and the Center for Cosmology and Particle Physics, 
    Department of Physics, New York University, New York, NY 10003, USA
    }%
  \thanks{E-mail: y.chan.12@ucl.ac.uk (J.~Y.~H.~Chan); 
  		          boris.leistedt.11@ucl.ac.uk (B.~Leistedt); 
		          t.kitching@ucl.ac.uk (T.~D.~Kitching); 
			  jason.mcewen@ucl.ac.uk (J.~D.~McEwen)}}

\markboth{IEEE Transcations on Signal Processing,~Vol.~65, No.~1, JAN 2017}%
{Chan \MakeLowercase{\textit{et al.}}: Second-generation Curvelets on the sphere}
%



\maketitle

\begin{abstract}
Curvelets are efficient to represent highly anisotropic signal content, such as a local linear and curvilinear structure. 
First-generation curvelets on the sphere, however, suffered from blocking artefacts. 
We present a new second-generation curvelet transform, where 
scale-discretised curvelets are constructed directly on the sphere. 
Scale-discretised curvelets exhibit a parabolic scaling relation,  
are well-localised in both spatial and harmonic domains, 
support the exact analysis and synthesis of both scalar and spin signals, 
and are free of blocking artefacts. 
We present fast algorithms to compute the exact curvelet transform, reducing 
computational complexity from $\mathcal{O}(L^5)$ to $\mathcal{O}(L^3\log_{2}{L})$ 
for signals band-limited at $L$. 
The implementation of these algorithms is made publicly available. 
Finally, we present an illustrative application demonstrating the effectiveness of curvelets 
for representing directional curve-like features in natural spherical images. 
\end{abstract}


%
\IEEEpeerreviewmaketitle

\section{Introduction}

\IEEEPARstart{S}{pherical} wavelets 
(\eg\ \cite{antoine:1999, antoine:1998,  mcewen:szip, narcowich:2006, baldi:2009, marinucci:2008, sanz:2006, 
 wiaux:2005, mcewen:2006:cswt2, wiaux:2007:sdw, leistedt:s2let_axisym, mcewen:2013:waveletsxv, 
 mcewen:s2let_localisation, mcewen:s2let_spin, mcewen:s2let_ridgelets, mcewen:2008:fsi, 
 michailovich:2010a, starck:2006,  geller:2008, geller:2010:sw, geller:2010}) 
 have proved to be highly successful in applications to 
cosmology (\eg\ \cite{vielva:2004, 
  mcewen:2008:ng, mcewen:2006:bianchi, 
  delabrouille:2009, vielva:2005,  
  vielva:2006, 
  lan:2008,
  mcewen:2007:isw2,  pietrobon:2006, 
  planck2013-p06, planck2013-p09, 
  planck2013-p20, planck2015-p4, 
  bobin:2013};
for a review, see \cite{mcewen:2006:review}), 
astrophysics (\eg\ \cite{schmitt:2012}), 
planetary science (\eg\ \cite{Holschneider:2003, audet:2014}), 
geophysics (\eg\ \cite{schmidt:2006, simons:2011}), 
and many other disciplines such as 
neuro-science (\eg\ \cite{Rathi:2011tq}). 
The reasons for this success are two-fold. 
Firstly, spherical data is common in nature, as demonstrated by the diverse range of applications above. 
In such cases, signals are naturally defined on the sphere and the most efficient analysis techniques are those that respect this underlying geometry. 
Secondly, different physical processes manifest on different scales, 
thereby imprinting scale-dependent, localised features within signals. 
Wavelets enable simultaneous extraction of both spectral and spatial information, 
thus making them a powerful analysis tool. 

In addition to scale-dependent and localised characteristics of signals, 
signals often contain directional and geometrical features, 
such as linear or curvilinear structures in 2D images (\eg\ edges), or 
sheet-like and filamentary structures in 3D space.  
Extraction of these features can in turn provide insightful information 
about the origin of signals or play crucial roles in diagnostic uses. 
Traditional wavelets, however, fall short of capturing this signal content effectively, 
which has motivated the development of a variety of directional or geometric wavelets. 

Ridgelet  (\eg\ \cite{candes:1999:ridgelets}) 
and curvelet  (\eg\ \cite{candes:1999curvelets, Candes:2004, Candes:2005bg, Candes:2005bg2}) 
transforms are of substantial interest since they provide 
efficient representations of line-type structures and exploit the anisotropic contents of signals. 
Among the two, ridgelets are limited to application to signals with global straight-line features only. 
In order to analyse local linear or curvilinear structures, which are dominant in nature, 
a block ridgelet-based transform, namely the first-generation curvelet transform, has been proposed. 
In Euclidean space, such a curvelet transform consists of applying an isotropic wavelet transform, 
followed by a special partitioning of the image and the application of the ridgelet transform to 
local overlapping blocks \cite{candes:1999curvelets}. 
The overlapping blocks, which are used to mitigate blocking artefacts, increase the redundancy, 
hence increasing the computational storage and timing costs. 
The same authors proposed second-generation Euclidean curvelets, rectifying these issues,  
where the discrete frequency domain is tiled and a ridgelet transform is no longer required \cite{Candes:2004, Candes:2005bg, Candes:2005bg2}. 
The second-generation curvelet construction is conceptually more natural and enables faster algorithms,  
thus opening up a wider and more successful applicability of curvelets, particularly in the fields of image processing, seismic image recovery and scientific computing (for reviews of the planar ridgelet and curvelet transforms, see \cite{Ma:2009tj, Fadili:2012ht}). 

Recently, a new generation of ridgelets on the sphere has been constructed 
in \cite{mcewen:s2let_ridgelets}, which is applicable to study antipodal signals on the sphere and is capable 
to handle both scalar and spin signals. 
First-generation curvelets have also been constructed on the sphere \cite{Starck:2005hl}, 
where the \healpix \:\cite{gorski:2005} scheme of partitioning of the sphere is employed and 
a discrete planar ridgelet transform is performed on each block independently. 
First-generation spherical curvelets are therefore not defined natively on the sphere (but rather by stitching together planar patches).  Furthermore, unlike the approach of the first-generation planar curvelets, 
the twelve base-resolution faces of the \healpix \:pixelisation do not overlap. 
This unavoidably leads to blocking artefacts \cite{Starck:2005hl}. In addition, in this framework curvelets larger than the scale of the base-resolution face cannot be constructed and first-generation spherical curvelets only satisfy the typical curvelet parabolic scaling relation (\ie\ ${\rm width} \approx {\rm length}^2$) in the Euclidean limit.

In this article, we construct a second-generation curvelet transform, which is \emph{not} built on a ridgelet transform, following a similar motivation to the development of the second-generation planar curvelets. Second-generation curvelets live natively on the sphere (\ie\ are not reliant on a specific pixelisation such as \healpix), 
are free from any blocking artefacts, 
satisfy the typical curvelet parabolic scaling relation, and 
support the exact synthesis of a band-limited signal from its curvelet coefficients 
(\ie\ capture all of the information content of the signal of interest without loss). 

It is possible to construct spherical curvelets through the inverse stereographic projection of planar wavelets \cite{antoine:1998,antoine:1999,wiaux:2005}, 
but the continuous scales required for the continuous analysis prevents exact reconstruction of the signals in practice. 
We therefore construct scale-discretised curvelets using the general spin scale-discretised wavelet framework presented in \cite{mcewen:s2let_spin}, where the dilations of the curvelets are directly defined in harmonic space and exact synthesis can be performed in practice. This framework also enables a straightforward generalisation of the curvelet transform to spin signals, where the spin value of the curvelets is a free parameter.
Depending on the desired applications, different spin values can be chosen: 
spin-0 for analysing scalar signals, spin-1 for vector fields and spin-2 for polarisation studies, for example. 
Furthermore, we show explicitly how the parabolic scaling relation is rendered 
in (spin) spherical polar coordinates by 
setting the absolute value of the azimuthal frequency index of spin spherical harmonic functions equal to the angular frequency index. 
Curvelets constructed in this manner exhibit many desirable properties, as listed earlier, which are lacking in alternative constructions (\eg\ \cite{Starck:2005hl}).

Having constructed scale-discretised curvelets applicable to transform signals of arbitrary spin on the sphere, 
we then present a fast algorithm to compute the curvelet transform exactly and efficiently. 
Optimisation is achieved by working in a rotated coordinate system that renders the harmonic representation of many curvelets coefficients zero; only relatively small number of non-zero terms need then be computed.  
This fast algorithm leverages novel sampling theorems on the sphere \cite{driscoll:1994, mcewen:fssht} 
and on the rotation group \cite{mcewen:so3}, where the later is further optimised for curvelets. 
These curvelet algorithms are implemented in the existing \stwoletcode \:code \cite{leistedt:s2let_axisym} -- 
an implementation of the scale-discretised wavelet transform on the sphere -- 
and are made publicly available. 

The remainder of this paper is organised as follows. 
In Section \ref{sec:1}, we construct curvelets that live natively on 
the sphere. 
The properties of curvelets and their differences to axisymmetric and directional wavelets on the sphere are highlighted. 
In Section \ref{sec:computation}, we derive exact and efficient algorithms for the numerical implementation of the curvelet transform. 
Numerical accuracy and computational-time scaling for a complete forward and inverse transform are evaluated. 
In Section \ref{sec:3}, we present an illustrative application, where a spherical image of a natural scene is analysed and 
the performance of curvelets and directional wavelets is compared. 
We conclude in Section \ref{sec:4}, outlining possible applications of the spherical curvelet transform 
and propose future extensions of this work. 
  
\section{Scale-discretised Curvelets on the Sphere}\label{sec:1}
In this section we present mathematical preliminaries of spin spherical harmonics and the rotation group, 
followed by the second-generation curvelet construction, properties and transform. 
We construct curvelets that are defined natively on the sphere, 
exhibit the standard curvelet parabolic scaling relation, 
are well-localised in both spatial and harmonic domains, 
and support the exact analysis and synthesis of both scalar and spin signals.
The construction follows closely to that of spin scale-discretised wavelets \cite{mcewen:s2let_spin}, 
and their analogous scalar forms \cite{wiaux:2007:sdw, leistedt:s2let_axisym, mcewen:2013:waveletsxv, mcewen:s2let_localisation}, 
except that the directionality component of curvelets is designed differently. 
For further details of the scale-discretised wavelet framework, 
including a review of harmonic analysis of (spin) functions on the sphere, 
we refer interested readers to the aforementioned papers. 
We conclude this section by noting some properties of curvelets, 
comparing those to axisymmetric and directional wavelets. 

Throughout this paper, all the transforms are formulated for the general spin setting, 
where the scalar setting can be simply rendered by setting the spin value $s \in \mathbb{Z}$ to zero. 
Also, we consider signals on the sphere band-limited at $\elmax$ throughout, 
\ie\ $\shc{\fs}{\el}{\m} =0$, $\forall \el\geq\elmax$, 
where $\shc{\fs}{\el}{\m}$, with interger $\el,\m \in\integers$, $|\m| \leq\el$, 
are the spin spherical harmonic coefficients of a spin signal of interest $\fs \in \ltwo(\sphere)$,  
and are given by the usual projection onto each spin spherical harmonic (basis) function $\sshf{\el}{\m}{\spin} \in \ltwo(\sphere)$: 
$\fslm =\innerp{\fs}{\sshf{\el}{\m}{\spin}}$. 

\subsection{Spin Spherical Harmonics And The Rotation Group}\label{subsec:rotate}
Curvelets probe signal content not only in scale and position on the sphere, but also in orientation. 
As such, the resulting curvelet coefficients live on the rotation group \mbox{$\ltwo(\sothree)$}. 
The Wigner $\dmatbig$-functions $\Dlmn \in \ltwo(\sothree)$, with natural $\el\in\naturals$
and integers $\m,\n\in\integers$, $|\m|,|\n|\leq\el$, are the matrix
elements of the irreducible unitary representation of the rotation
group \sothree\, \cite{varshalovich:1989}.  
Consequently, they or their conjugate\footnote{
The Wigner $\dmatbig$-functions satisfy the conjugate symmetry
relation
$\dmatbig_{\m\n}^{\el\cconj}(\eul) = (-1)^{\m+\n}
\dmatbig_{-\m,-\n}^{\el}(\eul)$.
} \Dlmnc\ form a complete set of orthogonal bases in $\ltwo(\sothree)$.
%
The Wigner \dmatbig-functions may also be related to the spin-$s$ spherical harmonics by 
\begin{equation}
  \label{eqn:ssh_wigner}
  \sshfarg{\el}{\m}{\sas}{\spin} = (-1)^\spin 
  \sqrt{\frac{2\el+1}{4\pi} } \:
  \dmatbig_{\m(-\spin)}^{\el\:\cconj}(\sab,\saa  ,0)
  \spcend , 
\end{equation} 
\cite{goldberg:1967}, 
for $\spin\in\integers$, $\el\in\naturals$ and $\m\in\integers$ such that 
$|\m|\leq\el$, $|\spin|\leq\el$, 
and where the Wigner $\dmatbig$-functions are parameterised by the Euler angles $\rho = (\alpha, \beta, \gamma)$, 
where $\alpha \in [0, 2\pi)$, $\beta \in [0,\pi]$ and $\gamma \in [0, 2\pi)$. 
Here we adopt the $zyz$ Euler convention, which corresponds to the rotation of a
physical body in a fixed coordinate system about the $z$, $y$ and $z$ axes 
by $\eulc$, $\eulb$ and $\eula$, respectively. 
$\dmatbig_{\m\n}^{\el}(\rho)$ can be further decomposed by 
\begin{equation}  
\dmatbig_{\m\n}^{\el}(\euls)
  =\exp{-\img\m\eula}\: 
   \dmatsmall^{\el}_{\m{n}} (\eulb) \:
    \exp{-\img\n\eulc}
   \spcend   , 
\end{equation}
where the Wigner small-$d$-functions may be expressed as 
\begin{eqnarray}
  \label{eqn:wigner_sum_reln}
  \dlmnb  =(-1)^{\el-\n} 
  \sqrt{(\el+\m)! (\el-\m)! (\el+\n)! (\el-n)!} \nonumber \\
  \times
 \sum_{k} 
 \frac{(-1)^{k}\: \Big(\sin{\frac{\beta}{2}} \Big)^{2\el-m-n-2k} \: \Big(\cos{\frac{\beta}{2}} \Big)^{m+n+2k} }
 {k! (\el-\m-k)! (\el-\n-k)! (\m+\n+k)!}
  \spcend , 
\end{eqnarray}
in which the sum is performed over all values of $k$ such that the arguments of the factorials are non-negative. 
It follows that the spin spherical harmonics can be expressed 
in spherical coordinates $\omega = (\theta, \varphi)$, with 
colatitude $\saa \in [0,\pi]$ and longitude $\sab \in [0,2\pi)$, by\footnote{
Note that we adopt the Condon-Shortley phase
convention such that the conjugate symmetry relation
\mbox{$\sshfargc{\el}{\m}{\omega}{\spin} = (-1)^{\spin+\m}
  \sshfargsp{\el}{-\m}{\omega}{-\spin}$} holds.}
\begin{equation}  
\label{eqn:spin_sphericalHarmonics}
  \sshfarg{\el}{\m}{\sa}{\spin} 
  = (-1)^s \sqrt{\frac{2\el+1}{4\pi}}\: 
    \dmatsmall^\el_{\m{(-\spin)}}(\theta)\:
    \exp{\img\m\varphi}
 \spcend . 
\end{equation}

\subsection{Curvelet Construction}\label{subsec:curveletConstruct}
We construct scale-discretised curvelets {$\swav^{(\wscale)} \in \ltwo(\sphere)$} in harmonic space in factorised form 
\begin{equation}
  \label{eqn:wav_factorized}
  \sshc{\wav}{\el}{\m}{\spin}^{(\wscale)} \equiv 
  \sqrt{\frac{2\el+1}{8\pi^2}} \:
  \wavker^{(\wscale)}(\el) \: \sshc{\wavsteer}{\el}{\m}{\spin}
  \spcend , 
\end{equation} 
where $\sshc{\wav}{\el}{\m}{\spin}^{(\wscale)} =
\innerp{\swav^{(\wscale)}}{\sshf{\el}{\m}{\spin}}$ 
are the spin spherical harmonic coefficients of the curvelets 
with $\sshf{\el}{\m}{\spin} \in \ltwo(\sphere)$ denoting the spin spherical harmonic functions,  
for $\spin\in\integers$, $\el\in\naturals$ and $\m\in\integers$ such that $|\m|\leq\el$, $|\spin|\leq\el$. 
The angular localisation of the $j$-th scale curvelet is characterised by the kernel $\wavker^{(\wscale)} \in \ltwo(\reals^{+})$ 
whose construction follows exactly the same as that of the spin directional scale-discretised wavelets 
given in \cite{mcewen:s2let_spin}. 
On the other hand, 
the directional localisation of curvelets is controlled by the directional component ${}_\spin\wavsteer$, 
with harmonic components $\sshc{\wavsteer}{\el}{\m}{\spin} = \innerp{{}_\spin\wavsteer}{\sshf{\el}{\m}{\spin}}$. 
It is this directional component that is defined in a way such that the parabolic scaling relation typical of curvelets is satisfied. 

We now show that the standard curvelet parabolic scaling relation can be rendered in spherical coordinates 
by considering spin spherical harmonics with the absolute value of the azimuthal frequency index equal to the angular frequency index, 
\ie\ $|\m|=\el$. 
Specifically, we show that the full-width-half-maximum (FWHM) of the colatitude $\saa \in [0,\pi]$ part of ${}_{\spin}Y_{\el \el}$ is approximately the square of that of the longitude $\sab \in [0,2\pi)$ part. 
Such a parabolic scaling also applies to curvelets since their harmonic coefficients are constucted from a windowed sum of spherical harmonics, 
with a central dominant angular frequency. 

We define the FWHM, which characterises the width about the peak of a function, as the 
difference between $\theta$ (or $\varphi$) at which (the real or imaginary part of) 
the function ${}_{\spin}Y_{\el \el}$ is equal to half of its maximum value. 
It is then straightforward to show that 
\begin{eqnarray}
\label{eq:FWHMphi}
{\rm FWHM}_{\varphi}= 
2{\varphi_0} = \frac{2}{\el}\cos^{-1}\Big(\frac{1}{2}\Big) = \frac{2\pi}{3\el}   
  \spcend, 
\end{eqnarray}
where ${\varphi_0}$ is the angle at the half maximum of the $\varphi$-part of ${}_{\spin}Y_{\el \el}$ (\ie\ real or imaginary part of $\exp{\img\el\varphi}$) within the interval $0 < {\varphi_0} <\pi/2 $. 
The $\theta$-dependence of ${}_{\spin}Y_{\el \el}$ is determined by the Wigner small-$d$-function   
\begin{eqnarray}
\label{eq:dll_s}
d^\el_{\el \: -s}(\theta)= (-1)^{\el+s} \sqrt{\frac{(2\el)!}{(\el-s)!(\el+s)!}}
\sin^{\el+s}{\frac{\theta}{2}}
\cos^{\el-s}{\frac{\theta}{2}}
  \spcend, 
\end{eqnarray} 
which attains its maximum at 
\begin{eqnarray}
\label{eq:theta_peak}
\theta_{\rm max} = \cos ^{ - 1}\Big({\frac{-s}{\el}}\Big)
  \spcend, 
\end{eqnarray} 
for $|\spin|\leq\el$.  As $s$ varies from $0$ to $\el$, $\theta_{\rm max}$ takes the value from $\pi/2$ to $\pi$ (indicating a change of the colatitude position at which spin-curvelets are centred, as will become explicit in the complete curvelet construction that follows). 
Furthermore, note that for the scalar setting $s=0$ and also for $s \ll \el$, 
\eqn{\ref{eq:dll_s}} reduces to the form of $A_{\el}\sin^{\el}{\theta}$, 
where $A_{\el}$ is a function of $\el$ contributing to the overall magnitude of ${}_{\spin}Y_{\el \el}$ only (and thus can be ignored in the evaluation of ${\rm FWHM}_{\theta}$). It follows that 
\begin{align}
	{\rm FWHM}_{\theta}= 2\Big(\frac{\pi}{2}-{\theta_0}\Big) =\pi -2 \sin^{-1}\Big(\frac{1}{2^{1/\el}}\Big)  
	  \spcend, 
\label{eq:FWHMtheta}
\end{align}
where  ${\theta_0}$ is the angle at the half maximum of the $\theta$-part of ${}_{\spin}Y_{\el \el}$ within the interval of $0 < {\theta_0} <\pi/2$. 
\eqn{\ref{eq:FWHMtheta}} can be further rearranged to 
\begin{align}
          \sin \Big(\frac{\pi-{\rm FWHM}_{\theta}}{2} \Big) =  2^{-u}    
	  \spcend, 
\label{eq:FWHMtheta2}
\end{align}
where $u = 1/\el$. In the limit $\el \rightarrow \infty$, $u$ and ${\rm FWHM}_{\theta}$ both approach to zero. 
Hence, by taking Taylor's expansion at both sides of \eqn{\ref{eq:FWHMtheta2}} one obtains 
\begin{eqnarray}
1-\frac{1}{8}\,{\rm FWHM}^{2}_{\theta} 
\approx 1- (\ln 2)\frac{1}{\el} 
  \spcend, 
\end{eqnarray}
which implies the important curvelet parabolic scaling relation 
\begin{eqnarray}
\label{eq:parabolic}
{\rm FWHM}^{2}_{\theta} 
\approx {\rm FWHM}_{\varphi}  
 \spcend .
\end{eqnarray}
We stress that the cases of spin value $s=0$ or $s \ll \el$, for which the parabolic scaling relation has been shown to hold, are common in real-life applications since physical signals are often scalar or have a low spin value. 
Furthermore, low-$\el$ information is often not probed by curvelets but rather by a scaling function, which will be discussed subsequently. 
Nevertheless, for completeness and clarity, 
we remark that in extreme cases when $s =\el$, the approximate parabolic scaling relation still holds, 
with the value of ${\rm FWHM}^{2}_{\theta}$ double that in the scalar setting. 
We also note that the parabolic scaling relation may start to deteriorate as $s \rightarrow (\el -1)$ due to the asymmetry of $d^\el_{\el \: -s}(\theta)$ about $\theta_{\rm max}$. However, for at least $s \simeq \floor{\el/2}$ (a very conservative limit), empirical numerical findings show that any deviation from the scalar setting is insignificant, so the parabolic scaling relation remains to hold. We refer readers to Appendix A for further details. 

Apart from setting $|\m| =\el$, the directionality component of curvelets 
$\sshc{\wavsteer}{\el}{\m}{\spin}$, without loss of generality, is defined to satisfy the condition 
\begin{equation}
  \label{eqn:directionality_normalisation}
  \summ \vert \sshc{\wavsteer}{\el}{\m}{\spin} \vert^2 = 1
  \spcend, 
\end{equation}
for all values of $\el$ for which $\sshc{\wavsteer}{\el}{\m}{\spin}$ are non-zero for at least one value of $m$. 
Consequently, the directionality component reads 
\begin{eqnarray}
\sshc{\widetilde{\wavsteer}}{\el}{\m}{\spin} =  \frac{1}{\sqrt 2}
  \Biggl\{ \begin{array}{ll} 
  \ 
  {(-1)^{\m}}\:{\delta_{\el\m}} , & m< 0\\ 
  \  
   \delta_{\el\m}, &  m \geq 0 \spcend, \end{array} 
    \spcend, 
\end{eqnarray} 
for all $\el$ of interest (with largest possible domain $0 < \el < L$) and $|\m| < L$.  
Here, $\delta$ denotes the Kronecker delta function, 
and the symbol $\tilde{\cdot}$ denotes that the quantity is associated to 
unrotated curvelets offset from the North pole (see \eqn{\ref{eq:theta_peak}}). 
It is desirable to centre curvelets on the North pole, so that the Euler angles parameterising curvelet coefficients have their standard interpretation, and their directionality component are given by the harmonic rotation 
\begin{eqnarray}
\sshc{\wavsteer}{\el}{\m}{\spin} = \sum_{\n=-\el}^{\el}  \Dlmn(\rho^{\star}) \sshc{\widetilde{\wavsteer}}{\el}{\n}{\spin} , 
\end{eqnarray} 
where $\rho^{\star}$ is the Euler angle describing the rotation to the North pole and is specified subsequently. 

Following \cite{mcewen:s2let_spin, wiaux:2007:sdw, leistedt:s2let_axisym, mcewen:2013:waveletsxv, mcewen:s2let_localisation}, the scale-discretised curvelet kernel for scale $j$ is constructed by 
\begin{equation}
  \wavker^{(\wscale)}(\el) \equiv  \wavker_\dilparam(\dilparam^{-\wscale} \el)
  \spcend,
\end{equation} 
which is generated from $\wavker_\dilparam(t) \equiv \sqrt{ k_\dilparam(\dilparam^{-1} t) - k_\dilparam(t) }$,  
has a compact support on
$\el \in \bigl[\floor{\dilparam^{\wscale-1}},
\ceil{\dilparam^{\wscale+1}} \bigr]$,
with $\floor{\cdot}$ and $\ceil{\cdot}$ denoting the floor and ceiling functions respectively,  
and reaches a peak of unity at $\el = \dilparam^{\wscale}$. 
The functions 
$k_\dilparam$ is defined by 
\begin{equation}
  k_\dilparam(t) \equiv 
  \frac{\int_{t}^1\frac{{\rm d}t^\prime}{t^\prime}  \:
   s_\dilparam^2(t^\prime)} 
   {\int_{\dilparam^{-1}}^1\frac{{\rm d}t^\prime}{t^\prime}\:
   s_\dilparam^2(t^\prime)}
    \spcend,
\end{equation}
which is unity for $t < \dilparam^{-1}$, zero for $t >1$, and is smoothly
decreasing from unity to zero for $t \in [\dilparam^{-1},1]$. 
It is defined through the inifinitely differentiable Schwartz function 
\begin{equation}
  s_\dilparam(t) 
  \equiv s\biggl( \frac{2\dilparam}{\dilparam-1} (t-\dilparam^{-1})-1\biggr)
  \spcend,
\end{equation}
where
\begin{equation}
  s(t) \equiv \Biggl\{ \begin{array}{ll} 
  \ 
  \exp{-(1-t^2)^{-1}}, & t\in[-1,1] \\ \  0, & t \notin [-1,1]\end{array} 
  \spcend, 
\end{equation} 
which has compact support $t \in[\dilparam^{-1}, 1]$, 
for dilation parameter $\dilparam \in \realsnz$, $\dilparam>1$.
Note that $\lambda =2$ corresponds to a common dyadic transform (\ie\ the mother curvelet is dilated by powers of two). 

As noted earlier, without applying any rotation the constructed spin-$s$ curvelets 
${}_{s}\widetilde{\wav}^{(j)}_{\el \m}$ are not centred on the North pole but at colatitude 
\begin{equation}
\label{eq:theta_peakj}
  \vartheta^{j} =  \cos ^{ - 1}\Big({\frac{-s}{\lambda^{j}}}\Big)
  \spcend, 
\end{equation}
\cf\, \eqn{\ref{eq:theta_peak}}, 
which lies in the range [$\pi/2$, $\pi$]. Explicitly, spin-$0$ curvelets are centred along the equator ($-x$-axis), and for higher-$s$ curvelets up to $s=\el$, curvelets effectively move down to be centred around the South pole. 
Curvelets are therefore rotated to the North pole by a rotation with Euler angle $\rho^{\star} =(0, \vartheta^{j}, 0)$. 

Scaling functions $\swavs \in \ltwo(\sphere)$, which are required to 
probe the low-frequency content (approximation-information) of the signal not probed by curvelets, 
are defined explicitly in \cite{mcewen:s2let_spin, leistedt:s2let_axisym, mcewen:2013:waveletsxv, mcewen:s2let_localisation}. 
They are chosen to be axisymmetric since directional structure of the approximation-information of signal is typically not of interest. 
For completeness, we repeat their definition here, which reads 
\begin{equation}
  \sshc{\wavs}{\el}{\m}{\spin} 
  \equiv 
  \sqrt{\frac{2\el+1}{4\pi}}
  \sqrt{ k_\dilparam(\dilparam^{-\wscalemin} \el)} \: \kron{\m}{0}
  \spcend   , 
\end{equation}
where $J_{0}$ is the minimum scale to be probed by curvelets.

\subsection{Curvelet Tiling And Properties}
\begin{figure}  
     \centering
     \subfigure[${}_0\wav^{(\wscale=1)}$]{\includegraphics[width=0.30\columnwidth,  trim = 3.87cm 14cm 22cm 1.62cm, clip=true]{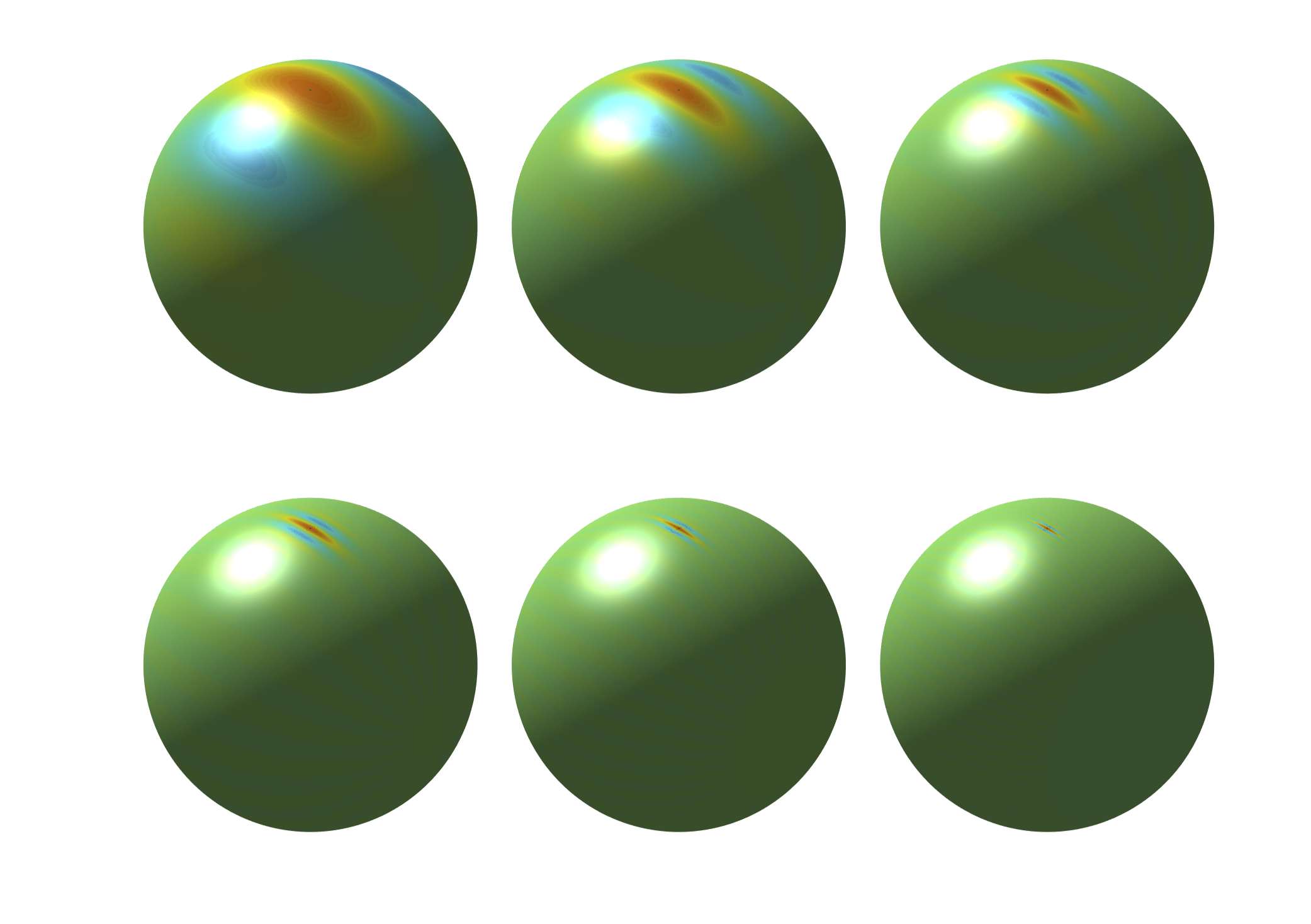}} 
     \subfigure[${}_0\wav^{(\wscale=2)}$]{\includegraphics[width=0.305\columnwidth,  trim = 13.3cm 14cm 12.51cm 1.62cm, clip=true]{s2let_demo7_Spin0_N256_L256_B2_Jmin2_cur_jet_notitle}}
     \subfigure[${}_0\wav^{(\wscale=3)}$]{\includegraphics[width=0.32\columnwidth,  trim = 22.87cm 14cm 2.5cm 1.62cm, clip=true]{s2let_demo7_Spin0_N256_L256_B2_Jmin2_cur_jet_notitle}}\\ 
     \subfigure[${}_0\wav^{(\wscale=4)}$]{\includegraphics[width=0.30\columnwidth,  trim = 3.87cm 2.5cm 22cm 13.3cm, clip=true]{s2let_demo7_Spin0_N256_L256_B2_Jmin2_cur_jet_notitle}} 
     \subfigure[${}_0\wav^{(\wscale=5)}$]{\includegraphics[width=0.305\columnwidth,  trim = 13.3cm 2.5cm 12.51cm 13.3cm, clip=true]{s2let_demo7_Spin0_N256_L256_B2_Jmin2_cur_jet_notitle}} 
          \subfigure[${}_0\wav^{(\wscale=6)}$]{\includegraphics[width=0.32\columnwidth,  trim = 22.87cm 2.5cm 2.5cm 13.3cm, clip=true]{s2let_demo7_Spin0_N256_L256_B2_Jmin2_cur_jet_notitle}}      
      \caption{Scalar scale-discretised curvelets on the sphere ($L=256,  \lambda=2$). Cuvelets are rotated to be centred on the North pole. Notice that the characteristic curvelet parabolic scaling (\ie\ ${\rm width} \approx {\rm length}^2$) makes them highly anisotropic and directionally sensitive.}
   \label{fig:constructedcurvelets}
\end{figure}

\begin{figure}  
     \centering
     \subfigure[${\mathbb R}{\rm e}\bigl\{{}_2\wav^{(\wscale=1)}\bigr\}$]{\includegraphics[width=0.30\columnwidth,  trim = 3.87cm 14cm 22cm 1.62cm, clip=true,]{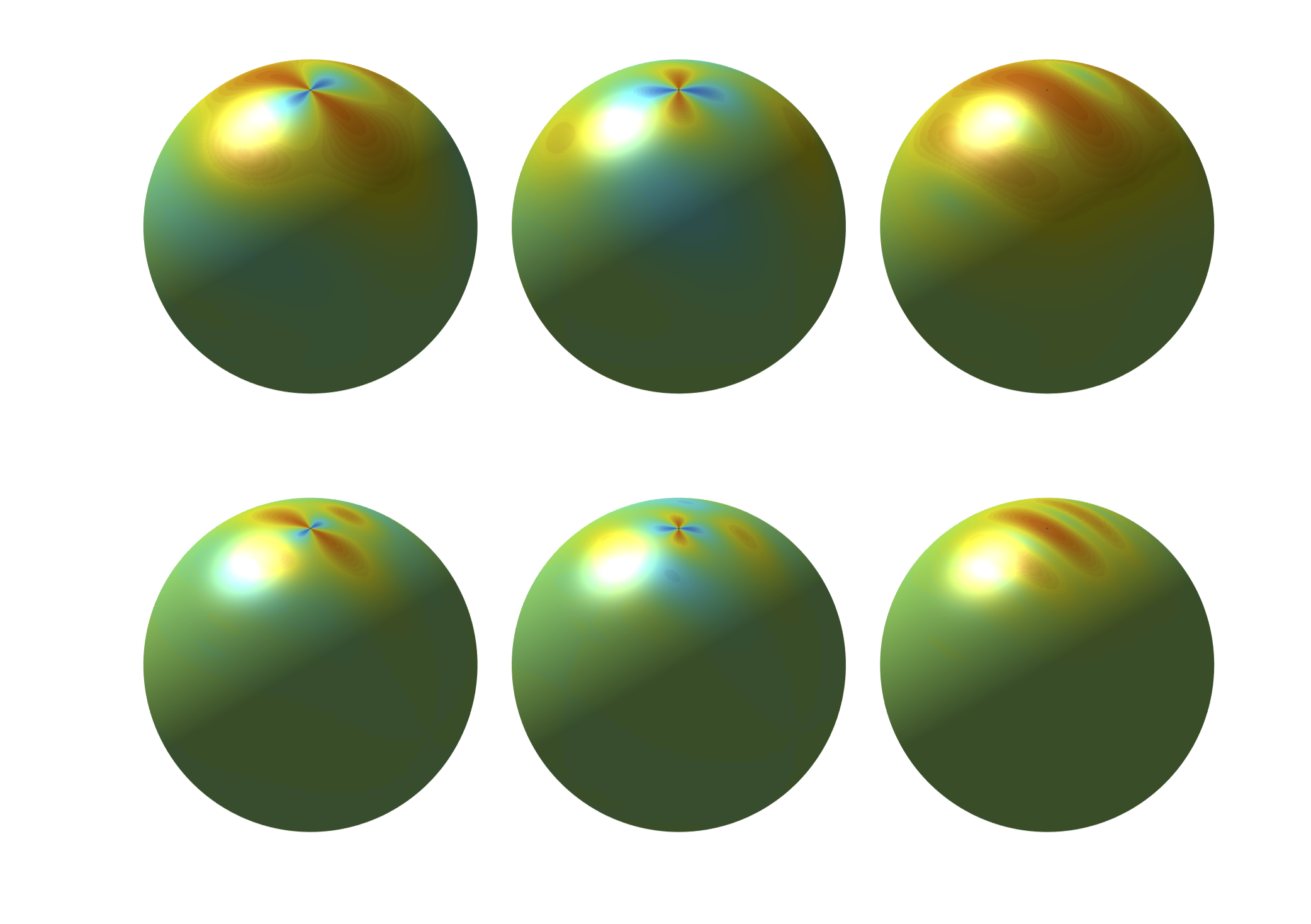}}
     \subfigure[${\mathbb I}{\rm m}\bigl\{{}_2\wav^{(\wscale=1)}\bigr\}$]{\includegraphics[width=0.305\columnwidth,  trim = 13.3cm 14cm 12.51cm 1.62cm, clip=true,]{s2let_demo7_Spin2_N256_L256_B2_Jmin2_cur_jet_notitle}}
     \subfigure[Abs$\bigl\{{}_2\wav^{(\wscale=1)}\bigr\}$]{\includegraphics[width=0.32\columnwidth,  trim = 22.87cm 14cm 2.5cm 1.62cm, clip=true,]{s2let_demo7_Spin2_N256_L256_B2_Jmin2_cur_jet_notitle}}\\ 
     \subfigure[${\mathbb R}{\rm e}\bigl\{{}_2\wav^{(\wscale=2)}\bigr\}$]{\includegraphics[width=0.30\columnwidth,  trim = 3.87cm 2.5cm 22cm 13.3cm, clip=true,]{s2let_demo7_Spin2_N256_L256_B2_Jmin2_cur_jet_notitle}}
     \subfigure[${\mathbb I}{\rm m}\bigl\{{}_2\wav^{(\wscale=2)}\bigr\}$]{\includegraphics[width=0.305\columnwidth,  trim = 13.3cm 2.5cm 12.51cm 13.3cm, clip=true,]{s2let_demo7_Spin2_N256_L256_B2_Jmin2_cur_jet_notitle}}
          \subfigure[Abs$\bigl\{{}_2\wav^{(\wscale=2)}\bigr\}$]{\includegraphics[width=0.32\columnwidth,  trim = 22.87cm 2.5cm 2.5cm 13.3cm, clip=true,]{s2let_demo7_Spin2_N256_L256_B2_Jmin2_cur_jet_notitle}}     
      \caption{Spin-2 scale-discretised curvelets on the sphere ($L=256,  \lambda=2$). Cuvelets are rotated to be centred on the North pole. Notice that the absolute value of the spin-curvelet is directional and the characteristic curvelet parabolic scaling (\ie\ ${\rm width} \approx {\rm length}^2$) makes them highly anisotropic and directionally sensitive.}
   \label{fig:spin_curvelets}
\end{figure}

Examples of  spin-0 (scalar) and spin-2 curvelets rotated to the North pole of the sphere are plotted in Figures \ref{fig:constructedcurvelets} and \ref{fig:spin_curvelets}, respectively. 
We highlight that the spin value is a free parameter, 
allowing easy construction of curvelets of any spin $\spin\in\integers$. 
Note also that as the scale $j$ increases, curvelets become increasingly elongated and exhibit increasingly higher directional sensitivity and anisotropic features (for spin curvelets, notice their absolute value is directional). 
This feature, originated by the satisfaction of the parabolic scaling relation, is absent in directional scale-discretised wavelets. 

 
The harmonic tiling of scale-discretised curvelets is schematically depicted in 
the right-most panel of Figure \ref{fig:ml_tiling}, 
along with the tilings of the the axisymmetric scale-discretised wavelets \cite{leistedt:s2let_axisym}\,(left-most panel) 
and the directional scale-discretised wavelets \cite{mcewen:s2let_spin, mcewen:2013:waveletsxv, mcewen:s2let_localisation}\,(middle panel) 
for comparison purposes. 
Axisymmetric wavelets probe signals in scale and position, but not in orientation, by tiling the line $\m = 0$ only. 
Directional wavelets are capable of probing the directional features of signals, 
but do not exploit the geometric properties of structures in signals. 
Tiling therefore occurs up to a low azimuthal band-limit $N < L$ (typically only even or odd $\m$ are non-zero to enforce various azimuthal symmetries). 
In contrast to axisymmetric and directional wavelets, 
curvelets probe not only spatial and spectral information, 
but also both directional and geometric contents of a signal. 
Such an ability is afforded by their specific design to render the parabolic scaling relation. 
This standard curvelet scaling relation is imposed by the tiling of curvelets along the corresponding lines. 

\tikzset{%
solid/.style={dash pattern=},
dotted/.style={dash pattern=on \pgflinewidth off 2pt},
densely dotted/.style={dash pattern=on \pgflinewidth off 1pt},
loosely dotted/.style={dash pattern=on \pgflinewidth off 4pt},
dashed/.style={dash pattern=on 6pt off 3pt},
densely dashed/.style={dash pattern=on 6pt off 2pt},
loosely dashed/.style={dash pattern=on 6pt off 6pt},
dashdotted/.style={dash pattern=on 5pt off 1pt on \the\pgflinewidth off 1pt},
densely dashdotted/.style={dash pattern=on 3pt off 1pt on \the\pgflinewidth off 1pt},
loosely dashdotted/.style={dash pattern=on 8pt off 4pt on \the\pgflinewidth off 4pt},
}%
\tikzset{%
ultra thin/.style={line width=0.1pt},
very thin/.style={line width=0.2pt},
thin/.style={line width=0.4pt},
semithick/.style={line width=0.6pt},
thick/.style={line width=0.8pt},
very thick/.style={line width=1.2pt},
ultra thick/.style={line width=1.6pt},
extra thick/.style={line width=2.2pt},
exxtra thick/.style={line width=3.4pt}
}%
\tikzset{%
 font={\fontsize{10pt}{12}\selectfont}}
\pgfmathdeclarefunction{gauss}{2}{%
\pgfmathparse{1/(#2*sqrt(2*pi))*exp(-((x-#1)^2)/(2*#2^2))}}

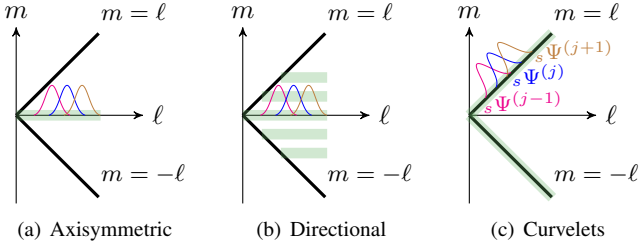
\begin{figure}
\centering
\subfigure[Axisymmetric]
{
\begin{tikzpicture}[scale=0.835, 
    axis/.style={thin, ->, >=stealth'},
    important line/.style={very thick},
    dashed line/.style={dashed, thin},
    pile/.style={thick, ->, >=stealth', shorten <=2pt, shorten
    >=2pt},
    every node/.style={color=black}
    ]
    \draw[axis] (-0.0,0)  -- (2,0) node(xline)[right]{\el};
    \draw[axis] (0,-1.4) -- (0,1.4) node(yline)[above] {\m};   
    \draw[important line] (0,0) coordinate (a_1) -- (1.3,1.3) coordinate (a_2);
    \draw[important line] (0,0) coordinate (b_1) -- (1.3,-1.3) coordinate (b_2);
    \coordinate (positiveline) at (1.85, 0.95);
    \node [above=1em] at (positiveline) {\small $\m =\el$};
    \coordinate (negativeline) at (2, -1.65);
    \node [above=1em] at (negativeline) {\small $\m = -\el$};
    	\draw[line width=4pt, color=green!50!black!60!white,opacity=0.3] (0,0) --+(0:1.33);
    \begin{scope}[rotate=0,scale=0.1,yshift=-3ex,xshift=14ex]
	\begin{axis}[every axis plot post/.append style={mark=none,domain=-2:2,samples=50,smooth},
			axis x line=none, axis y line=none,]
	\addplot[color=magenta] {gauss(0,0.75)};
	\end{axis}
\end{scope}
\begin{scope}[rotate=0,scale=0.1,yshift=-3ex,xshift=29ex]
	\begin{axis}[every axis plot post/.append style={mark=none,domain=-2:2,samples=50,smooth},
			axis x line=none, axis y line=none,]
	\addplot[color=blue] {gauss(0,0.75)};
	\end{axis}
\end{scope}
\begin{scope}[rotate=0,scale=0.1,yshift=-3ex,xshift=44ex]
	\begin{axis}[every axis plot post/.append style={mark=none,domain=-2:2,samples=50,smooth},
			axis x line=none, axis y line=none,]
	\addplot[color=brown] {gauss(0,0.75)};
	\end{axis}
\end{scope}
\end{tikzpicture}
}
%
\subfigure[Directional]
{
\begin{tikzpicture}[scale=0.835, 
    axis/.style={thin, ->, >=stealth'},
    important line/.style={very thick},
    dashed line/.style={dashed, thin},
    pile/.style={thick, ->, >=stealth', shorten <=2pt, shorten
    >=2pt},
    every node/.style={color=black}
    ]
    \draw[axis] (-0.0,0)  -- (2,0) node(xline)[right]{\el};
    \draw[axis] (0,-1.4) -- (0,1.4) node(yline)[above] {\m};   
    \draw[important line] (0,0) coordinate (a_1) -- (1.3,1.3) coordinate (a_2);
    \draw[important line] (0,0) coordinate (b_1) -- (1.3,-1.3) coordinate (b_2);
    \coordinate (positiveline) at (1.85, 0.95);
    \node [above=1em] at (positiveline) {\small $\m =\el$};
    \coordinate (negativeline) at (2, -1.65);
    \node [above=1em] at (negativeline) {\small $\m = -\el$};
    	\draw[line width=4pt, color=green!50!black!60!white,opacity=0.3] (0,0) --+(0:1.33);
	\draw[line width=4pt, color=green!50!black!60!white,opacity=0.3] (0.3,0.3) --+(0:1.03);
	\draw[line width=4pt, color=green!50!black!60!white,opacity=0.3] (0.6,0.6) --+(0:0.73);		
	\draw[line width=4pt, color=green!50!black!60!white,opacity=0.3] (0.3,-0.3) --+(0:1.03);
	\draw[line width=4pt, color=green!50!black!60!white,opacity=0.3] (0.6,-0.6) --+(0:0.73);		
    \begin{scope}[rotate=0,scale=0.1,yshift=-3ex,xshift=14ex]
	\begin{axis}[every axis plot post/.append style={mark=none,domain=-2:2,samples=50,smooth},
			axis x line=none, axis y line=none,]
	\addplot[color=magenta] {gauss(0,0.75)};
	\end{axis}
\end{scope}
\begin{scope}[rotate=0,scale=0.1,yshift=-3ex,xshift=29ex]
	\begin{axis}[every axis plot post/.append style={mark=none,domain=-2:2,samples=50,smooth},
			axis x line=none, axis y line=none,]
	\addplot[color=blue] {gauss(0,0.75)};
	\end{axis}
\end{scope}
\begin{scope}[rotate=0,scale=0.1,yshift=-3ex,xshift=44ex]
	\begin{axis}[every axis plot post/.append style={mark=none,domain=-2:2,samples=50,smooth},
			axis x line=none, axis y line=none,]
	\addplot[color=brown] {gauss(0,0.75)};
	\end{axis}
\end{scope}
\end{tikzpicture}
}
%
\subfigure[Curvelets]
{
\begin{tikzpicture}[scale=0.835, 
    axis/.style={thin, ->, >=stealth'},
    important line/.style={very thick},
    dashed line/.style={dashed, thin},
    pile/.style={thick, ->, >=stealth', shorten <=2pt, shorten
    >=2pt},
    every node/.style={color=black}
    ]
    \draw[axis] (-0.0,0)  -- (2,0) node(xline)[right]{\el};
    \draw[axis] (0,-1.4) -- (0,1.4) node(yline)[above] {\m};   
    \draw[important line] (0,0) coordinate (a_1) -- (1.3,1.3) coordinate (a_2);
    \draw[important line] (0,0) coordinate (b_1) -- (1.3,-1.3) coordinate (b_2);
    \coordinate (positiveline) at (1.85, 0.95);
    \node [above=1em] at (positiveline) {\small $\m =\el$};
    \coordinate (negativeline) at (2, -1.65);
    \node [above=1em] at (negativeline) {\small $\m = -\el$};
    	\draw[line width=4pt, color=green!50!black!60!white,opacity=0.3] (0,0) --+(45:1.85);
	\draw[line width=4pt, color=green!50!black!60!white,opacity=0.3] (0,0) --+(-45:1.85);
    \begin{scope}[rotate=45,scale=0.1,yshift=-2ex,xshift=20ex]
	\begin{axis}[every axis plot post/.append style={mark=none,domain=-2:2,samples=50,smooth},
			axis x line=none, axis y line=none,]
	\addplot[color=magenta] {gauss(0,0.75)};
	\end{axis}
\end{scope}
\begin{scope}[rotate=45,scale=0.1,yshift=-2ex,xshift=35ex]
	\begin{axis}[every axis plot post/.append style={mark=none,domain=-2:2,samples=50,smooth},
			axis x line=none, axis y line=none,]
	\addplot[color=blue] {gauss(0,0.75)};
	\end{axis}
\end{scope}
\begin{scope}[rotate=45,scale=0.1,yshift=-2ex,xshift=50ex]
	\begin{axis}[every axis plot post/.append style={mark=none,domain=-2:2,samples=50,smooth},
			axis x line=none, axis y line=none,]
	\addplot[color=brown] {gauss(0,0.75)};
	\end{axis}
\end{scope}
    \coordinate (wav) at (1.1, -0.11);
    \node [above=1em, color =blue] at (wav) {\footnotesize $\swav^{(\wscale)}$};
        \coordinate (wav) at (1.65, 0.28);
    \node [above=1em, color=brown] at (wav) {\footnotesize $\swav^{(\wscale +1)}$};
        \coordinate (wav) at (0.85, -0.54);
    \node [above=1em, color=magenta] at (wav) {\footnotesize $\swav^{(\wscale -1)}$};
\end{tikzpicture}
}
\caption[]{Tiling of different types of wavelets. Notice in particular that by tiling curvelets along the diagonal, an approximate parabolic scaling relationship is imposed.}
\label{fig:ml_tiling}
\end{figure}

\subsection{Curvelet Transform}\label{sec:curvtransform}
The curvelet transform is built upon the 
spin scale-discretised wavelet framework presented in \cite{mcewen:s2let_spin} where 
the discrete nature of the analysis scales (\ie\ $\wscale \in \naturals_0$) allows 
the exact reconstruction of band-limited signals from their curvelet coefficients. 
This is ensured since the curvelets constructed as defined previously satisfy the 
admissibility condition
\begin{equation}
  \label{eqn:admissibility}
  \frac{4\pi}{2\el+1}   
  \bigl\vert \sshc{\wavs}{\el}{0}{\spin} \bigr\vert^2 + 
  \frac{8\pi^2}{2\el+1} 
  \sum_{\wscale=\wscalemin}^\wscalemax
  \sumn \bigl\vert \sshc{\wav}{\el}{\n}{\spin}^{(\wscale)} \bigr\vert^2 = 1 
  \spcend ,  
  \quad \forall\el 
  \spcend  , 
\end{equation} 
\cite[and reference therein]{mcewen:s2let_spin}. 
Here we define curvelet analysis and synthesis operations (\ie\ forward and inverse transform). 
Curvelets are rotated to the North pole in the following description. 

\subsubsection{Curvelet Analysis}
The scale-discretised curvelet transform of a function 
$\fs \in {\rm L}^2 ({\mathbb S}^2)$ 
is defined by 
its directional convolution with the curvelets $\swav^{(\wscale)} \in {\rm L}^2 ({\mathbb S}^2)$, 
where curvelet coefficients are given by 
\begin{align}
  \wcoeff^{\swav^{(\wscale)}}(\eul) 
  & \equiv \innerp{\fs}{\rotarg{\eul} \: \swav^{(\wscale)}} \nonumber \\
  &= \int_\sphere \dmu{\sa} \: \fs(\sa) \: (\rotarg{\eul} \: \swav^{(\wscale)})^\cconj(\sa)
  \label{eqn:wav_analysis}
  \spcend ,
\end{align}
where the rotation operator $\mathcal{R}_{\rho}$ is parameterised by 
the Euler angles $\eul = (\alpha, \beta, \gamma) \in {\rm SO}(3)$, 
with $\alpha \in [0, 2\pi)$, $\beta \in [0,\pi]$ and $\gamma \in [0, 2\pi)$.   
\eqn{\ref{eqn:wav_analysis}} may be re-written as
%
\begin{equation}
  \label{eqn:wav_analysis_harmonic}
  \wcoeff^{\swav^{(\wscale)}}(\eul)   
  = \sumlmn
  \sshc{\f}{\el}{\m}{\spin} \:
  \sshc{\wav}{\el}{\n}{\spin}^{(\wscale)\cconj} \:
  \Dlmnpc  
  \spcend,
\end{equation}
where $\fslm = \innerp{\fs}{\sshf{\el}{\m}{\spin}}$ and
$\sshc{\wav}{\el}{\m}{\spin}^{(\wscale)} =
\innerp{\swav^{(\wscale)}}{\sshf{\el}{\m}{\spin}}$
are the spin spherical harmonic coefficients of the function of interest and curvelet,
respectively.

The Wigner coefficients of the wavelet coefficients defined on SO(3) are given by 
$\wigc{\bigl(\wcoeff^{\swav^{(\wscale)}}\bigr)}{\el}{\m}{\n} =
\innerp{ \wcoeff^{\swav^{(\wscale)}}}{\Dlmnc}$, which can be reduced to
\begin{equation}
  \label{eqn:wav_harmonic_coeff}
  \wigc{\bigl(\wcoeff^{\swav^{(\wscale)}}\bigr)}{\el}{\m}{\n}
  = \frac{8 \pi^2}{2\el+1} \:
  \sshc{\f}{\el}{\m}{\spin} \:
  \sshc{\wav}{\el}{\n}{\spin}^{(\wscale)\cconj}
  \spcend  .
\end{equation} 
As such, the forward curvelet transform may be computed via an inverse Wigner transform. 

The low-frequency content of the signal is captured by the scaling coefficients $\swavs \in \ltwo(\sphere)$, 
which are given by the axisymmetric convolution 
\begin{eqnarray}
  \label{eqn:analysis_scaling}
  \wcoeff^{\swavs}(\sa) 
  &\equiv&  \innerp{\fs}{\rotarg{\sa}\:\swavs}\\
  &=& \int_\sphere \dmu{\sa\p} \:\fs(\sa\p) \:
  (\rotarg{\sa}\:\swavs)^\cconj(\sa\p)
  \spcend ,
\end{eqnarray}
where $\mathcal{R}_{\omega}=\mathcal{R}_{(\phi, \theta, 0)}$. 
Since the scaling function is, by design, axisymmetric, its harmonic coefficients are non-zero for $\m=0$ only:
$\sshc{\wavs}{\el}{0}{\spin} \kron{\m}{0} =
\innerp{\swavs}{\sshf{\el}{\m}{\spin}}$. 
Decomposing the scaling coefficients into their harmonic expansion yields
\begin{equation}
  \label{eqn:analysis_scaling_harmonic}
  \wcoeff^{\swavs}(\sa)   
  = \sumlm
  \sqrt{\frac{4 \pi}{2 \el + 1}} \:
  \sshc{\f}{\el}{\m}{\spin} \:
  \sshc{\wavs}{\el}{0}{\spin}^\cconj \:
  \sshfarg{\el}{\m}{\sa}{0}
  \spcend , 
\end{equation} 
whose spherical harmonic coefficient is simply given by 
\begin{equation}
  \label{eqn:swav_harmonic_coeff}
  \shc{\bigl(\wcoeff^{\swavs}\bigr)}{\el}{\m}
  = \innerp{\wcoeff^{\swavs}}{{}_{0}\shf{\el}{\m}} 
  = 
  \sqrt{\frac{4 \pi}{2 \el + 1}} \:
  \sshc{\f}{\el}{\m}{\spin} \:
  \sshc{\wavs}{\el}{0}{\spin}^\cconj
  \spcend .
\end{equation}

\subsubsection{Curvelet Synthesis}
Provided that the admissibility condition in \eqn{\ref{eqn:admissibility}} is satisfied, 
the signal $\fs$ can be reconstructed exactly from its curvelet and scaling coefficients by 
\begin{align}
  \label{eqn:wav_synthesis}
  \fs(\sa) 
  =& \int_\sphere \dmu{\sa\p} \:
  \scoeff^{\swavs}(\sa\p) \: (\rotarg{\sa\p} \: \swavs)(\sa)\nonumber\\
  &+
  \sum_{\wscale=\wscalemin}^\wscalemax \int_\sothree \deul{\eul} \:
  \wcoeff^{\swav^\wscale}(\eul) \: (\rotarg{\eul} \: \swav^\wscale)(\sa)
  \spcend ,
\end{align}
where the invariant measure on SO(3) is 
${\rm d} {\varrho}(\rho)=\sin{\beta} \: {\rm d} \alpha \: {\rm d} \beta\: {\rm d} \gamma$, 
and $J_{0}$ and $J$ are, respectively, the minimum and maximum analysis depths considered, \ie\ $\wscalemin \leq \wscale \leq \wscalemax$. 
\eqn{\ref{eqn:wav_synthesis}} may be re-written as
%
\begin{align}
  \label{eqn:wav_synthesis_harmonic}
  \fs(\sa) 
  =& 
  \sumlm \biggl[
  \sqrt{\frac{4 \pi}{2 \el + 1}} \:  
  \shc{\bigl(\wcoeff^{\swavs}\bigr)}{\el}{\m} \:
  \sshc{\wavs}{\el}{0}{\spin} \nonumber\\
  &+
  \sum_{\wscale=\wscalemin}^\wscalemax \sumn
  \wigc{\bigl(\wcoeff^{\swav^{(\wscale)}}\bigr)}{\el}{\m}{\n} \:
  \sshc{\wav}{\el}{\n}{\spin}^{(\wscale)}
   \biggr ] \sshfarg{\el}{\m}{\sa}{\spin}
  \spcend , 
\end{align}
\cite{mcewen:s2let_spin}.
As such, the inverse curvelet transform of \eqn{\ref{eqn:wav_synthesis}} may be computed via a forward Wigner transform.
 
\section{Exact and Efficient Computation}\label{sec:computation}
In this section, we devise a fast algorithm to compute the curvelet transform, 
which is theoretically exact by appealing to sampling theorems on the sphere \cite{driscoll:1994, mcewen:fssht} and rotation group \cite{mcewen:so3}. The computational complexity of the algorithm attains $\order(\elmax^3 \log_{2} \elmax)$, compared to the naive scaling of $\order(\elmax^5)$. We then discuss the implementation of this algorithm and evaluate its performances in terms of both numerical accuracy and computation time via simulations of random test signals on the sphere. 

\subsection{Fast Algorithm}\label{subsec:fastalgo}
Wigner transforms can be computed efficiently using the fast algorithm of \cite{mcewen:so3} which 
reduces the complexity from $\order(\elmax^6)$ to $\order(\elmax^4)$.  
For (steerable) directional wavelet transforms, 
for which the wavelets have an azimuthal band-limit $\nmax$, the complexity is reduced to $\order(\nmax \elmax^3)$ 
and since typically $\nmax \ll \elmax$, the overall complexity of $\order(\elmax^3)$ is recovered.  
For curvelets, there is however no azimuthal band-limit so fast Wigner transforms can only yield $\order(\elmax^4)$. 
Here we develop an algorithm that attains $\order(\elmax^3 \log_{2} \elmax)$. 
This is achieved by first rotating Wigner coefficients of curvelets coefficients (rather than the curvelets themselves) 
and by optimising the fast Wigner transform for curvelets. 
We present these algorithmic details next, followed by a description of 
additional optimisation that are exploited to further speed up our implementation.  

 \subsubsection{Rotating Wigner coefficients} \label{subsubsec:rotatingwigner}
As highlighted in \sectn{\ref{sec:curvtransform}}, 
curvelets are centred on the North pole in our scale-discretised curvelet transform 
so that the Euler angles parameterising curvelet coefficients have their standard interpretation. 
However, our directly constructed curvelet ${}_s{\widetilde \wav}$ 
is naturally centred at a different position. 
A rotation is therefore needed: either by rotating the curvelets directly or 
by rotating the Wigner coefficients of the curvelet coefficients. 
By exploiting the unrotated curvelet's property that ${}_{s}\widetilde{\wav}_{\el \n}={}_{s}\widetilde{\wav}_{\el \el} \delta_{\el \n}$ and hence 
 \begin{align}
   \label{eqn:curvelet_Wigner_property}
   \wigc{\bigl(\widetilde{\wcoeff}^{{}_{s}\widetilde{\wav}^{(\wscale)}}\bigr)}{\el}{\m}{\n} 
   = \wigc{\bigl(\widetilde{\wcoeff}^{{}_{s}\widetilde{\wav}^{(\wscale)}}\bigr)}{\el}{\m}{\n} \: \delta_{|\n| \el}
 \spcend  ,    
 \end{align}
\ie\ unrotated Wigner coefficients are non-zero for $|n| = \el$ only,  
we show in the following that rotating Wigner coefficients rather than curvelets leads to an additional optimisation. 

The rotation of the Wigner coefficients for the forward transform proceeds as follows. 
The Wigner coefficients of unrotated curvelets 
(offset from the North pole) can be computed by 
$\widetilde{\wcoeff}^{{}_s{\widetilde \wav}^{(\wscale)}}({\eul})  \equiv \innerp{\fs}{\rotarg{\eul} \: {}_s{\widetilde \wav}^{(\wscale)}}$, 
but we require 
\begin{align}
 \wcoeff^{\swav^{(\wscale)}}({\eul})  \equiv \innerp{\fs}{\rotarg{\eul} \: \swav^{(\wscale)}} 
 =  \innerp{\fs}{\rotarg{\eul} \rotarg{\eul^{\star}} \: {\widetilde \wav}^{(\wscale)}} 
  \spcend , 
\end{align}
where ${\swav^{(\wscale)}}({\eul}) = \rotarg{\eul^{\star}}{{}_s{\widetilde \wav}^{(\wscale)}}({\eul})$ 
denotes curvelets centred on the North pole, and $\eul^{\star} = (0, \vartheta^{j}, 0)$ 
is the Euler angle defining rotation to the North pole. It follows that 
\begin{align}
  \label{eqn:rotateWigner}
\wcoeff^{\swav^{(\wscale)}}({\eul}) 
=
{\widetilde \wcoeff}^{{}_s{\widetilde \wav}^{(\wscale)}}({\eul}{\p}) 
  \spcend , 
\end{align}
where $\eul{\p}$ describes the rotation formed by compositing the
rotations described by $\eul$ and $\eul^{\star}$, 
\ie\ $\rotarg{\eul\p} =  \rotarg{\eul} \: \rotarg{\eul^{\star}}$\ . 
\eqn{\ref{eqn:rotateWigner}} then can be computed by  
 \begin{align}
  \label{eqn:curvelet_Wigner}
   \wigc{\bigl(\wcoeff^{\swav^{(\wscale)}}\bigr)}{\el}{\m}{k}= &
   \sum_{\n} 
   \wigc{\bigl({\widetilde \wcoeff}^{{}_s{\widetilde \wav}^{(\wscale)}}\bigr)}{\el}{\m}{\n} \:
   \dmatbig_{k \n}^{\el}{}^{*}({\eul^{\star}})      \\ 
   =& 
    \wigc{\bigl({\widetilde \wcoeff}^{{}_s{\widetilde \wav}^{(\wscale)}}\bigr)}{\el}{\m}{\el} \: \dmatbig_{k \el}^{\el}{}^{*}({\eul^{\star}}) \nonumber \\
   & + \wigc{\bigl({\widetilde \wcoeff}^{{}_s{\widetilde \wav}^{(\wscale)}}\bigr)}{\el}{\m}{(-\el)} \: \dmatbig_{k (-\el)}^{\el}{}^{*}({\eul^{\star}})
   \spcend  , 
\end{align}
where we have exploited the additive property of the Wigner \dmatbig-functions and 
 \eqn{\ref{eqn:curvelet_Wigner_property}}. 

For the inverse transform, unrotated Wigner coefficients, which are non-zero for $|k| = \el$ only, are computed by 
 \begin{align}
  \label{eqn:curvelet_Wigner_inverse}
   \wigc{\bigl({\widetilde \wcoeff}^{\swav^{(\wscale)}}\bigr)}{\el}{\m}{k}
   = &
    \sum_{n=-\el}^\el 
       \wigc{\bigl(\wcoeff^{\swav^{(\wscale)}}\bigr)}{\el}{\m}{\n} \: \dmatbig_{k \n}^{\el}{}^{*}({\eul^{\star^\prime}}) 
   \spcend  , 
\end{align}
where inverse rotation described by the Euler angle ${\eul^{\star^\prime}}=(0, -\vartheta^{j}, 0)$ is performed, 
\ie\ $\rotarg{\eul^{\star^\prime}} = \rotarg{\eul^{\star}}^{-1}$. 


Notice from \eqn{\ref{eqn:curvelet_Wigner}} and \eqn{\ref{eqn:curvelet_Wigner_inverse}} that the 
computational complexity of the rotation is $\order(\elmax^3)$ only. 
In contrast, if one choose to rotate curvelets directly, non-zero rotated curvelets coefficients will span across the domain of $\n < \elmax$, prohibiting additional optimisation enabled by computing only the non-zero coefficients. 

\subsubsection{Optimised Wigner Transform}\label{subsubsec:optimisedso3}
It is not possible to directly optimise the fast algorithm to compute Wigner transform presented in \cite{mcewen:so3} (where fast spin spherical harmonic transforms are used for intermediate calculations), even with minor modifications, since the order of summations needs to be altered. Instead, we adapt this approach by interchanging the order of summations and performing all computations explicitly.  

We adopt an equiangular sampling of the rotation group with sample positions given by 
\begin{equation}
  \label{eqn:nodes_alpha}
  \eulaiang = 
  \frac{2 \pi \eulai}{2{M}-1}, 
  \quad \mbox{where } \eulai \in \{ 0,1,\dotsc,2{M}-2 \}
  \spcend ,
\end{equation}
\begin{equation}
  \label{eqn:nodes_beta}
  \eulbiang = 
  \frac{\pi(2\eulbi+1)}{2\elmax-1}, 
  \quad \mbox{where } \eulbi \in \{ 0,1,\dotsc,\elmax-1 \}
  \spcend ,
\end{equation}
and
\begin{equation}
  \label{eqn:nodes_gamma}
  \eulciang = 
  \frac{2 \pi \eulci}{2\nmax-1}, 
  \quad \mbox{where } \eulci \in \{ 0,1,\dotsc,2\nmax-2 \}
  \spcend ,
\end{equation}
\cite{mcewen:so3}. 
The forward Wigner transform, optimised for curvelets, proceeds as follows. 
Firstly, a Fourier transform is performed over Euler angles $\eula$ and $\eulc$:
\begin{align}
  \label{eqn:Fmn}
  {\mathcal X}_{\m\n}(\eulbiang) 
   = & \sum_{\eulai=-(M-1)}^{M-1} \sum_{\eulci=-(\nmax-1)}^{\nmax-1} 
   {\widetilde \wcoeff}^{\swav^{(\wscale)}}(\eulaiang, \eulbiang, \eulciang)   \nonumber \\
    &  \times \exp{-\img(\m\eulaiang+ \n\eulciang)} /{(2M-1)(2\nmax-1)}
      \spcend ,
\end{align}
for $b \in \{ 0, \ldots, \elmax-1\}$, $|\m|, |n| \leq\el$, 
for which the computation can be reduced from $\order(\elmax^5)$ to $\order(\elmax^3 \log_{2} \elmax)$ 
by the fast Fourier transform (FFT), where $\order(\elmax) = \order(M) = \order(N)$.
Secondly, we employ the trick (\eg\ \cite{mcewen:fssht,mcewen:so3}) of 
extending ${{\mathcal X}}_{\m\n}(\eulbiang)$ to the domain $[0,2\pi)$ through the construction 
 \begin{equation}
   \label{eqn:TildaFmn}
{\overline{\mathcal X}_{\m\n}}(\eulbiang) =  
 \begin{cases}
   \: (-1)^{\m+\n}  {\mathcal X}_{\m\n}(-\eulbiang)  \: , & \eulbi \in \{ \elmax, ... 2\elmax-2\}   \\
   \: {\mathcal X}_{\m\n}(\eulbiang)  \: , & \eulbi \in \{ 0, ... \elmax-1\}  
          \spcend ,
 \end{cases}
\end{equation}
which can be computed for all arguments and indices in $\order(\elmax^3)$.  
Subsequently, it is possible to compute the Fourier transform of 
${\overline{\mathcal X}_{\m\n}}(\eulbiang)$ in $\eulb$:
 \begin{align}
   \label{eqn:Fmnmp}
  {\mathcal X}_{\m\n\m\p}
   = \frac{1}{(2\elmax-1)} 
    \sum_{b=-(\elmax-1)}^{\elmax-1} {\overline{\mathcal X}_{\m\n}}(\eulbiang)\: \exp{-\img\m\p\eulbiang} 
          \spcend ,
 \end{align}
which can be computed for all indices in $\order(\elmax^3 \log_{2} \elmax)$ by FFTs.
Thirdly, an exact quadrature for integration over $\eulb$ follows by 
\begin{equation}
 \label{eqn:gmnmp}
  {\mathcal Y}_{\m\n\m\p}
  =  (2 \pi)^2 
 \sum_{\m\pp=-(\elmax-1)}^{\elmax-1}
   {\mathcal X}_{\m\n\m\pp} \: 
   w(\m\pp - \m\p)
 \spcend ,
\end{equation}
where the weights are given by $ \weight(\m\p) = \int_0^\pi
\dx\eulb \sin\eulb \: \exp{\img \m\p \eulb}$, which can be evaluated
analytically \cite{mcewen:fssht}.  \eqn{\ref{eqn:gmnmp}} can be computed directly at $\order(\elmax^4)$ or through its dual Fourier representation in $\order(\elmax^3 \log_{2} \elmax)$ (noting it is essentially a discrete convolution; see \cite{mcewen:fssht}).
Finally, Wigner coefficients can be computed by 
 \begin{equation}
 \label{eqn:flml}
  \wigc{ \big( {\widetilde \wcoeff}^{\swav^{(\wscale)}} \big) }{\el}{\m}{\el}
  = \img^{\m-\el} 
  \sum_{\m\p=-(\elmax-1)}^{\elmax-1} 
    \dlmnhalfpi{\el}{\m\p}{\m} \:
      \dlmnhalfpi{\el}{\m\p}{\el} \:
       {\mathcal Y}_{\m\el\m\p}
 \spcend ,
\end{equation}
in $\order(\elmax^3)$, 
where \mbox{$\dlmnhalfpi{\el}{\m}{\n} \equiv \dlmn (\pi/2)$} for $|\m|, |\n| \leq \el$, and we have exploited \eqn{\ref{eqn:curvelet_Wigner_property}}.
The overall forward transform is dominated by the computation of \eqn{\ref{eqn:Fmn}} or \eqn{\ref{eqn:Fmnmp}} 
and thus scales as $\order(\elmax^3 \log_{2} \elmax)$.

The inverse Wigner transform, optimised for curvelets, proceeds as follows. 
Firstly, Fourier coefficients of the Wigner coefficients are computed by
\begin{align}
 \label{eqn:inverse_Fmnmp}
{\mathcal X}_{\m\n\m\p}
 = \img^{\n-\m} \:
  {\frac{2|\n|+1}{8\pi^2} } \:
    \dlmnhalfpi{|\n|}{\m\p}{\m} \:
      \dlmnhalfpi{|\n|}{\m\p}{\n} \:
       \wigc{\big( {\widetilde \wcoeff}^{\swav^{(\wscale)}} \big)}{|\n|}{\m}{\n}
 \spcend ,
\end{align}
in $\order(\elmax^3)$, 
where we have exploited \eqn{\ref{eqn:curvelet_Wigner_property}}.

Secondly, curvelet coefficients are computed from the Fourier representation of their Wigner representation by 
\begin{align}
  \label{eqn:inverse_feul}
    {\widetilde \wcoeff}^{\swav^{(\wscale)}}(\eulaiang, \eulbiang, \eulciang) 
    = & \sum_{\m=-(M-1)}^{M-1} \sum_{\n=-(\nmax-1)}^{\nmax-1} \sum_{\m\p=-(\elmax-1)}^{\elmax-1} {\mathcal X}_{\m\n\m\p}     \nonumber \\ 
    &  \times  \exp{\img(\m\eulaiang+ \m\p\eulbiang+ {n}\eulciang)}
\end{align}
for which the computation can be reduced from $\order(\elmax^6)$ to $\order(\elmax^3 \log_{2} \elmax)$ by FFTs. 
The samples of ${\widetilde \wcoeff}^{\swav^{(\wscale)}}$ computed over $\eulb \in (\pi, 2\pi)$ are discarded. The overall inverse transform is dominated by the computation of \eqn{\ref{eqn:inverse_feul}} and thus scales as $\order(\elmax^3 \log_{2} \elmax)$. 

\subsubsection{Additional optimisations}\label{subsubsec:optimise_options} 
We construct a multi-resolution algorithm 
exploiting the reduced band-limit $L_j = \lambda^{j+1}$ of the curvelets for scales $j< J-1$ 
such that the minimal number of samples on the sphere is used to reconstruct curvelet coefficients for each scale 
(see also \cite{leistedt:s2let_axisym, mcewen:s2let_spin}). 
Consequently, only the finest curvelet scales at the largest $j \in \{{J-1, J}\}$ are computed at maximal resolution (corresponding to the band-limit of the signal) and thereby dominate the computation. 
The overall complexity of computing both forward and inverse wavelet transforms, including all scales, is thus 
$\order(\elmax^3\log_{2} \elmax)$. 
In addition, for real signals, we exploit their conjugate symmetry which leads to a further reduction of computational and memory requirements by a factor of two. 

\subsection{Implementation}
We have implemented the spherical curvelet transform in 
the existing \stwoletcode \:code\footnote{\url{http://www.s2let.org}}, 
and the fast Wigner transform algorithm optimised for curvelets in the existing \sothreecode \:code\footnote{\url{http://www.sothree.org}}. 
The \stwoletcode\: package, which currently supports 
scale-discretised scalar and spin 
axisymmetric wavelet \cite{leistedt:s2let_axisym}, 
directional wavelet \cite{mcewen:2013:waveletsxv, mcewen:s2let_localisation, mcewen:s2let_spin} 
and ridgelet \cite{mcewen:s2let_ridgelets} transforms, 
relies on the 
\sshtcode \:code\footnote{\url{http://www.spinsht.org}} 
\cite{mcewen:fssht} to compute spherical harmonic transforms, the
\sothreecode \:code \cite{mcewen:so3} to compute Wigner transforms and the
\fftwcode \:code\footnote{\url{http://www.fftw.org}} \cite{FFTW05} to compute fast Fourier transforms. 
Its core algorithms are implemented in C, with also Matlab, Python, IDL and JAVA interfaces provided, and \healpix \:maps are also supported.

\subsection{Numerical Experiments}
 %
\begin{figure}
     \centering
     \subfigure[Maximum error\label{fig:numaccuracy}]{\includegraphics[width=1\columnwidth,  trim = 0.1cm 0.22cm 2.cm 0.5cm, clip=true]{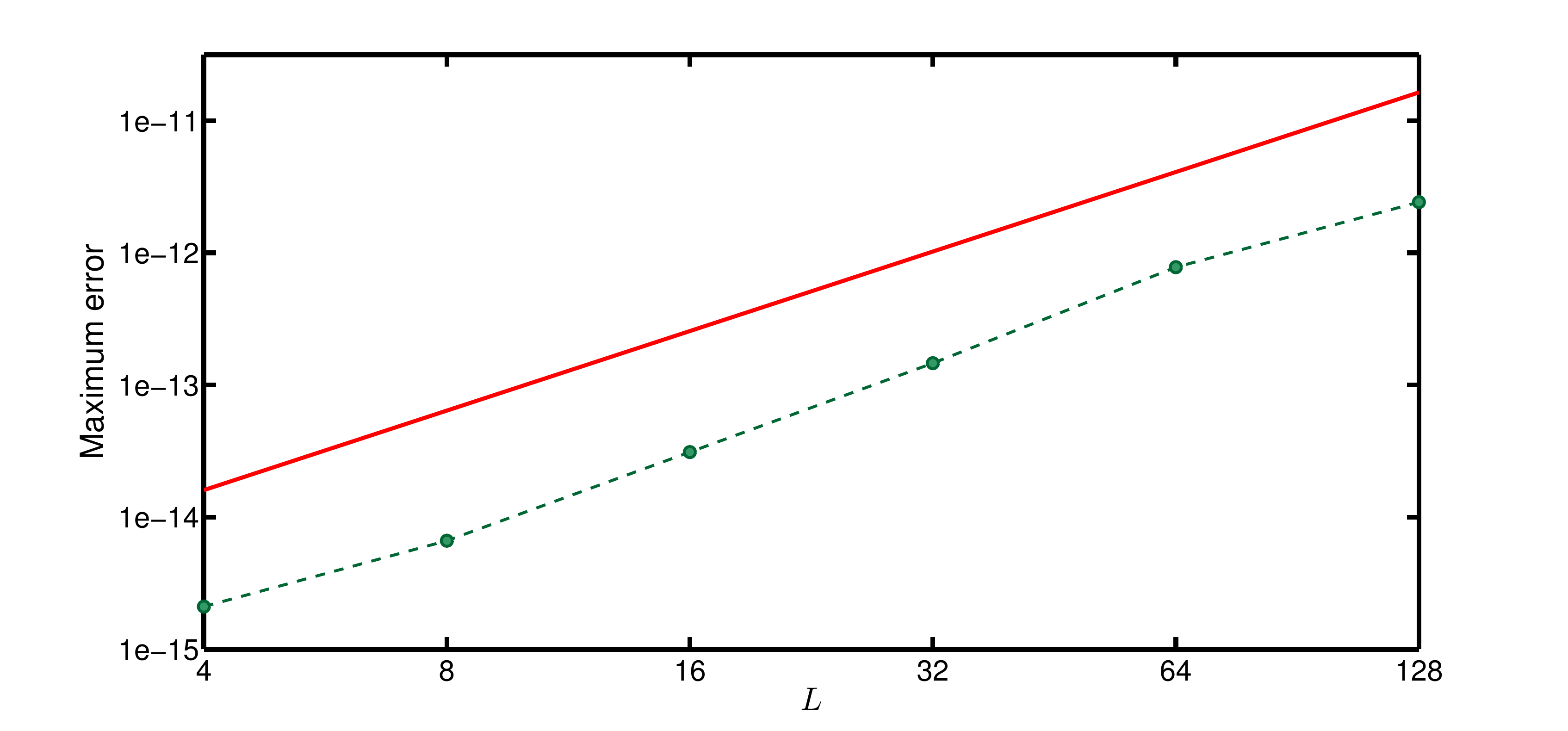}}\\ 
     \subfigure[Computation time \label{fig:timing}]{\includegraphics[width=1\columnwidth,  trim = 0.0cm 0.22cm 2cm 0.45cm, clip=true]{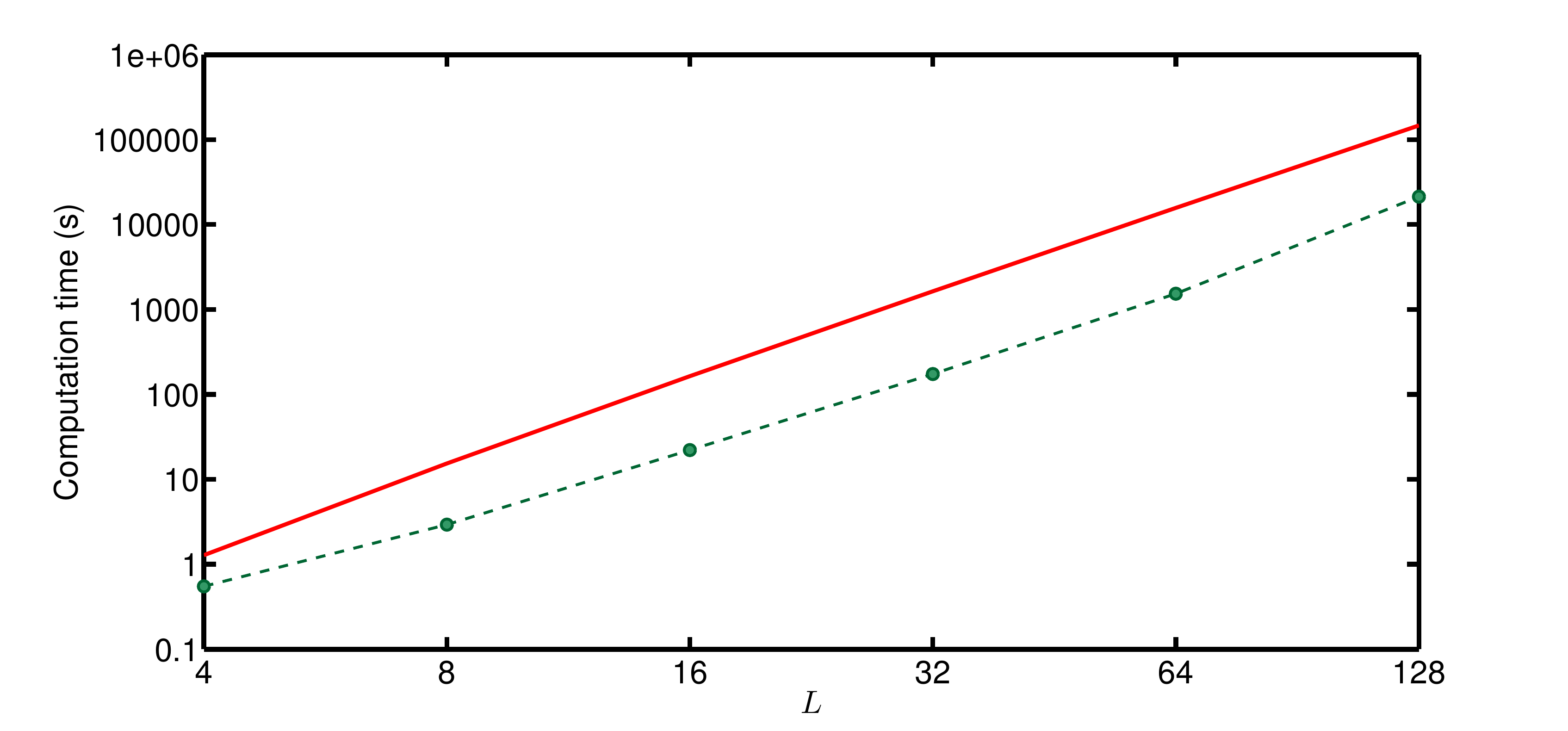}} 
     \vspace{-0.45cm}
      \caption{Numerical accuracy and computation time of the spherical curvelet transform, averaged over three round-trip transforms of random test signals. Numerical accuracy attains close to machine precision and is found empirically (shown by the dashed green line) to scale as $\mathcal{O}(L^2)$ (shown by the solid red line). Computation time is found empirically to scale closely to the theoretical prediction of  $\mathcal{O}(L^3\log_{2}{L})$. Plots showing the empirical results for spherical axisymmetric and directional wavelets can be found in \cite{leistedt:s2let_axisym, mcewen:s2let_spin} respectively.}
\end{figure}
 
We evaluate the numerical accuracy and computation time 
of the scale-discretised curvelet transform implemented in the \stwoletcode \:code. 
Random band-limited test signals are simulated on the sphere through the inverse spherical harmonic transform of 
their spherical harmonic coefficients $\fslm$ with real and imaginary parts uniformly distributed in the interval [-1,1]. 
We then perform a forward curvelet transform, followed by an inverse transform to reconstruct the original signals from their curvelet coefficients. 
Three repeats of the complete transform are performed for each $L$, where $L=\{4, 8, 16, 32, 64, 128\}$. 
All numerical experiments are performed on a workstation with 2$\times$12 core 1.8 GHz Intel(R) Xeon(R) processors and 256 GB of RAM (but parallelisation of the code has not been performed to fully exploit the multi-core architecture). 

 \subsubsection{Numerical Accuracy}
Numerical accuracy of a round-trip curvelet transform is
measured by the maximum absolute error between the spherical harmonic
coefficients of the original test signal $\fslm^{\rm o}$ and the
recomputed values $\fslm^{\rm r}$, \ie\
 $
  \epsilon = \mathop{\rm max}_{\el,\m} \:
  \bigl | \fslm^{\rm r} - \fslm^{\rm o} \bigr |
$.
Results of the numerical accuracy tests, averaged over three random test signals, 
are plotted in  \fig{\ref{fig:numaccuracy}}.  
We plot results for a scalar signal, although the accuracy for spin signals is identical since the spin number is simply a parameter of
the transform. The numerical accuracy of the round-trip
transform is close to machine precision and found empirically to scale
as $\order(\elmax^2)$.
%

 \subsubsection{Computation Time}
Computation time is measured by the round-trip computation time taken
to perform a forward and inverse curvelet transform.  Results of the
computation time tests, averaged over three random test signals, are 
plotted in \fig{\ref{fig:timing}}.  
We plot results for a scalar signal, although the computation time for signals of different spin numbers is identical. 
The computational complexity of the curvelet transform is found empirically to scale closely to the 
theoretical prediction of $\order(\elmax^3\log_{2}\elmax)$. 

\section{Illustration}\label{sec:3}
\begin{figure}
     \centering
     {\includegraphics[width=0.46\columnwidth,  trim = 0.cm 0cm 9.1cm 0.15cm, clip=true]{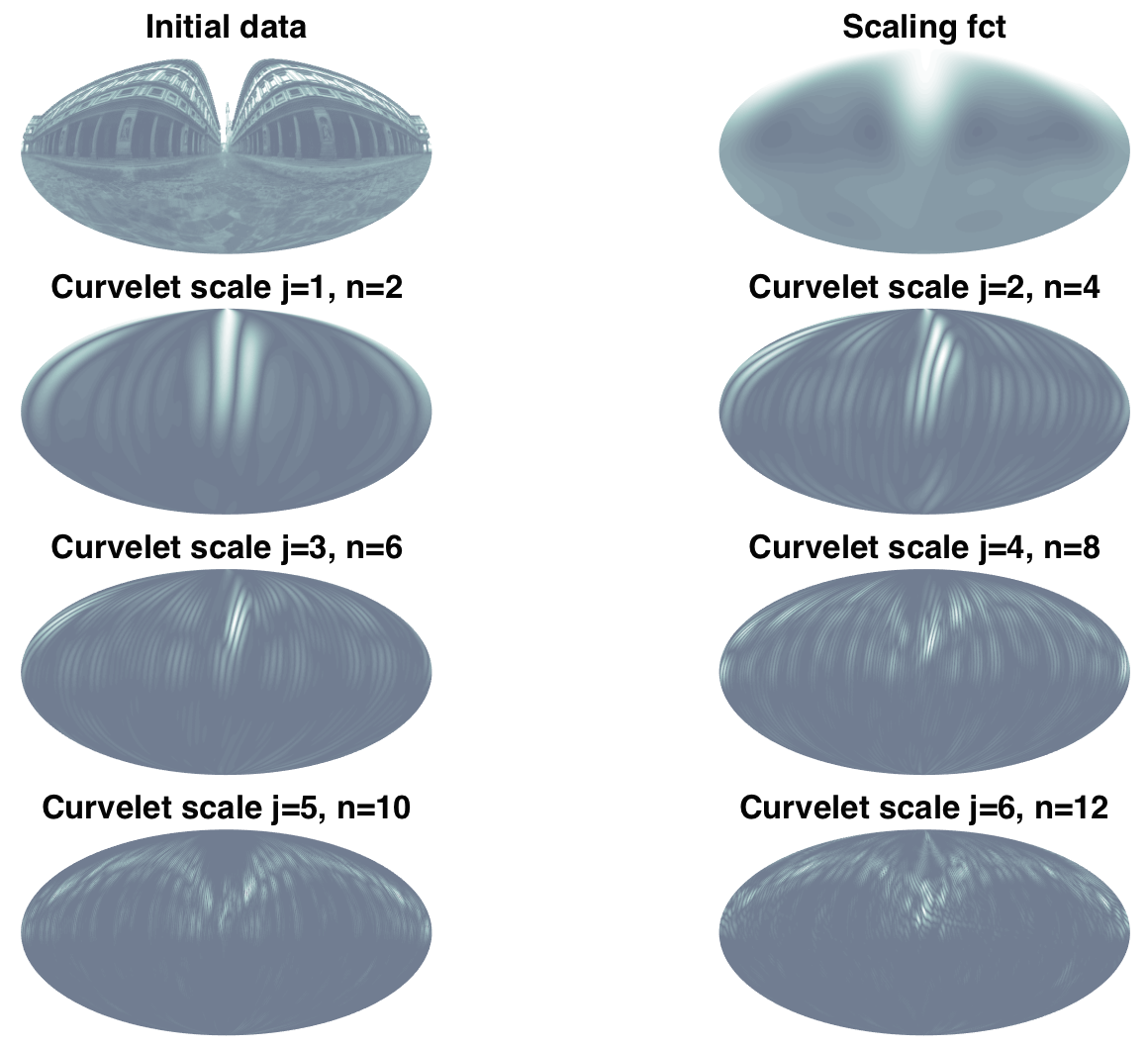}   
      \includegraphics[width=0.46\columnwidth,  trim = 9.cm 0cm 0.1cm 0.15cm, clip=true]{s2let_N256_L256_B2_Jmin3_curvelet_Uffiz}}
      \vspace{-0.2cm}
      \caption{Plots of scaling coefficients and curvelet coefficients obtained from the Uffizi Gallery at various scales and orientations. Notice the ability of curvelets to extract oriented, spatially localised, scale-dependent features in the light probe images and their very high sensitivity to line and curvilinear structures.}
      \label{fig:coef_lightprobe}
\end{figure}
\begin{figure}
     \centering
     \begin{tikzpicture}
        \node[anchor=south west,inner sep=0] at (-0.35,0){\includegraphics[width=0.89\columnwidth, height=5.5cm, trim = 1.4cm 1.cm 1.2cm 0.8cm, clip=true]{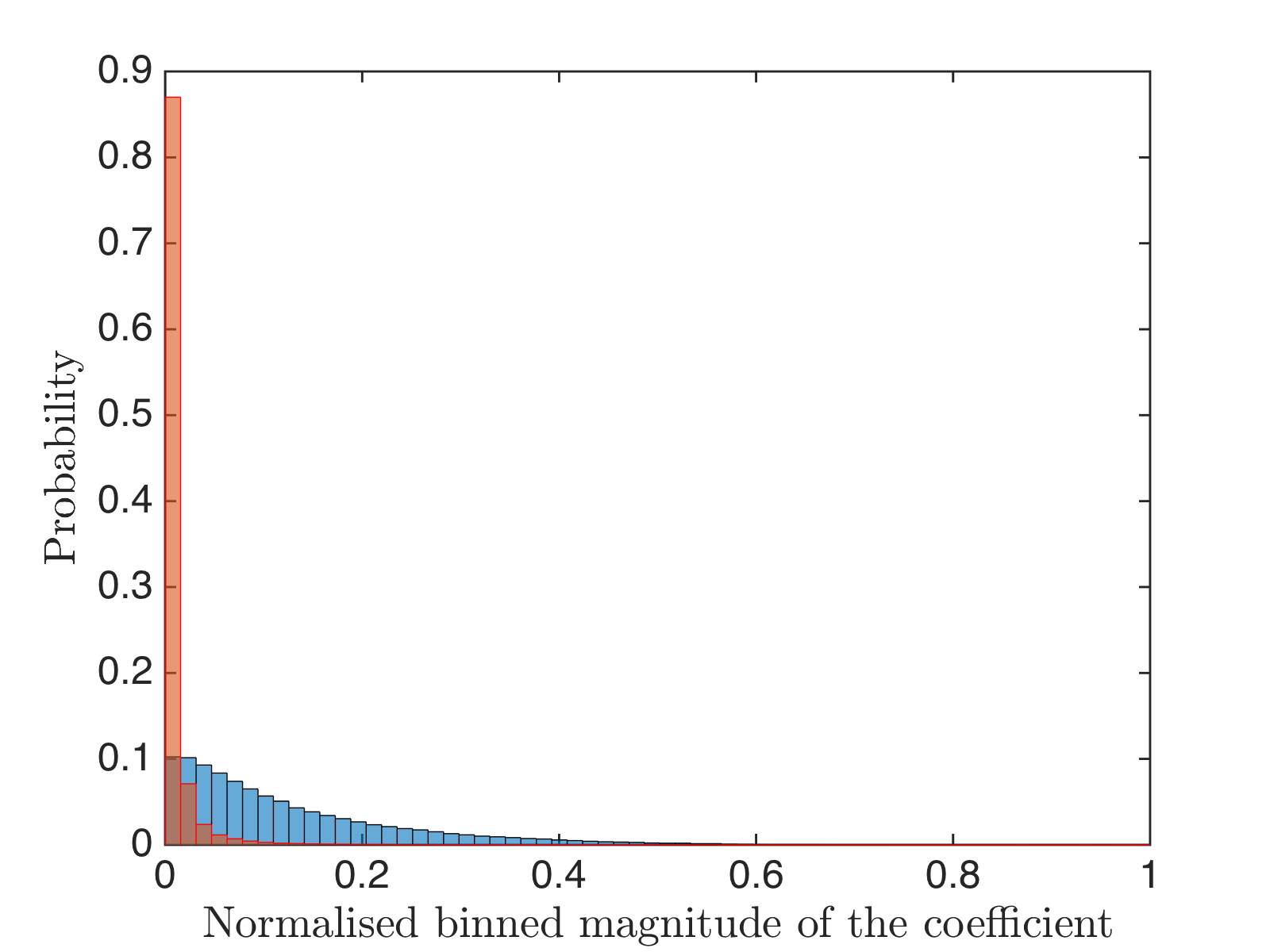}};
        \node[rotate=90, anchor=west, right] at (-0.55,2) {\small Probability} ;
        \node[anchor=west, right] at (2,-0.28) {\small Coefficient magnitude} ;
\end{tikzpicture}
\vspace{-0.3cm}
    \caption{Histogram showing the probability of the coefficients of wavelets (in blue) and curvelets (in red) obtained from the light probe image plotted in \fig{\ref{fig:coef_lightprobe}} for scale $j=6$ against the normalised coefficient magnitudes. The vertical axis shows the number of the wavelet or curvelet coefficients at each magnitude interval divided by the total count of the coefficients. The horizontal axis is normalised to unity by the maximum binned magnitude of the coefficients for comparison purposes. Curvelets yield a sparser representation than directional wavelets: there are many small curvelet coefficients and only few large coefficients.}
    \label{fig:histogram_lightprobe}
\end{figure}
In this section we present a simple application and analyse a spherical image of a natural scene with scale-discretised curvelets. 
We show that the spherical curvelet decomposition is sparse, with few large curvelet coefficients and many small coefficients. 
The ability of curvelets to represent natural spherical images efficiently is demonstrated using 
the light probe image of the Uffizi Gallery in Florence \cite{PaulDebevec:lightprobe}\footnote{\url{http://www.pauldebevec.com/Probes/}}, 
which contains substantial line and curvilinear structures. 
For simple illustrative purposes, the image is band-limited to $L = 256$ 
and the image intensity is clipped and rescaled before the curvelet transform is performed. 
Plots of the scaling coefficients and curvelet coefficients are shown in \fig{\ref{fig:coef_lightprobe}}. 
These plots show clearly that curvelets extract oriented, spatially localised, scale-dependent features in the image and 
are highly sensitive to edge-like features. 

To compare the performance of curvelets and directional wavelets,  
both transforms are applied with the same parameters and the directional wavelets have 
azimuthal limit set to $N=L$ for fair comparison with curvelets. 
We plot the histogram of curvelet and directional wavelet coefficients for scale $j =6$ 
of the Uffizi image in \fig{\ref{fig:histogram_lightprobe}}. 
It is apparent that curvelets yield a sparser representation than directional wavelets, 
where not only are there many small curvelet coefficients and few large coefficients, 
but the decay in number of coefficients is much faster than that of directional wavelets. 
This sparseness of curvelet representations of natural spherical image can be exploited in practical applications 
such as denoising, inpainting, and data compression, for example. 
%

\section{Conclusion}\label{sec:4}

In this article, we construct the second-generation curvelet transform that lives natively on the sphere. 
This curvelet transform exhibits the typical curvelet parabolic scaling relation for efficient representation of highly anisotropic signal content. 
It does not exhibit blocking artefacts due to special partitioning, 
does not rely on ridgelet transform, and 
admits exact inversion for signals defined on the sphere.  
Scale-discretised curvelets are constructed based on the general spin scale-discretised wavelet framework, 
which supports both scalar and spin settings. 
Fast algorithms to compute the exact forward and inverse curvelet transform are devised and 
are implemented in the existing \stwoletcode \:code package, which 
leverages a novel sampling theorem on the rotation group whose implementation is further optimised for curvelets. 
Through simulations, we demonstrate that the numerical accuracy of our transforms is close to machine precision 
and the computational speed scales as $\mathcal{O}(L^3\log_{2}{L})$, compared to the naive scaling of $\mathcal{O}(L^5)$. 
Our implementation of the curvelet transform is made publicly available. 

We illustrate the effectiveness of spherical curvelets for decomposing images 
with substantial line and curvilinear structures using an example natural spherical image. 
The curvelet decomposition is found to be sparser than the directional wavelet analysis in this example. 
This sparseness can be exploited in applications to data compression and signal processing (\eg\ to mitigate noise or handle incomplete data-set). 
More generally, the curvelets developed in this paper may find wide applications 
to transform scalar or spin signals acquired on the sphere 
where anisotropic and geometric structures in the signal content are of interest. 
For example, curvelets could be applied to identify and characterise (granules and) sunspots of the Sun and to study whole-sky polarisation signals, 
which are crucial in understanding cosmic magnetic fields ubiquitous in the Universe. 
%

In addition, for data sets where different signal characteristics are targeted at different scales, 
a hybrid use of curvelets and the other type of wavelets, where 
curvelets are tiled at some scales and the axisymmetric or directional wavelets are tiled at others, 
can be considered. 
This hybrid transform, or the curvelet transform, 
may also be extended to the three-dimensional ball (\ie\ solid sphere formed by augmenting the sphere with radial line) \cite{leistedt:flaglets, mcewen:flaglets_sampta}. 
Such tools could be used to study a diverse range of data-sets defined on the ball, 
such as cosmological 21-cm tomographic data, which is an important probe for understanding what happened when the first stars and first black holes formed and how the Universe transited from almost featureless to structure-filled as seen today. 
They are also important for weak gravitational lensing studies, in which the signal is a spin-2 field on the ball, that can be used to test the nature of dark energy and dark matter.

\appendices
\section{Parabolic scaling of spin curvelets}
\begin{figure}
     \centering
\begin{tikzpicture}
        \node[anchor=south west,inner sep=0] at (-0.35,0) {\includegraphics[width=0.95\columnwidth, trim = 0cm 0.cm 0cm 0.58cm, clip=true]{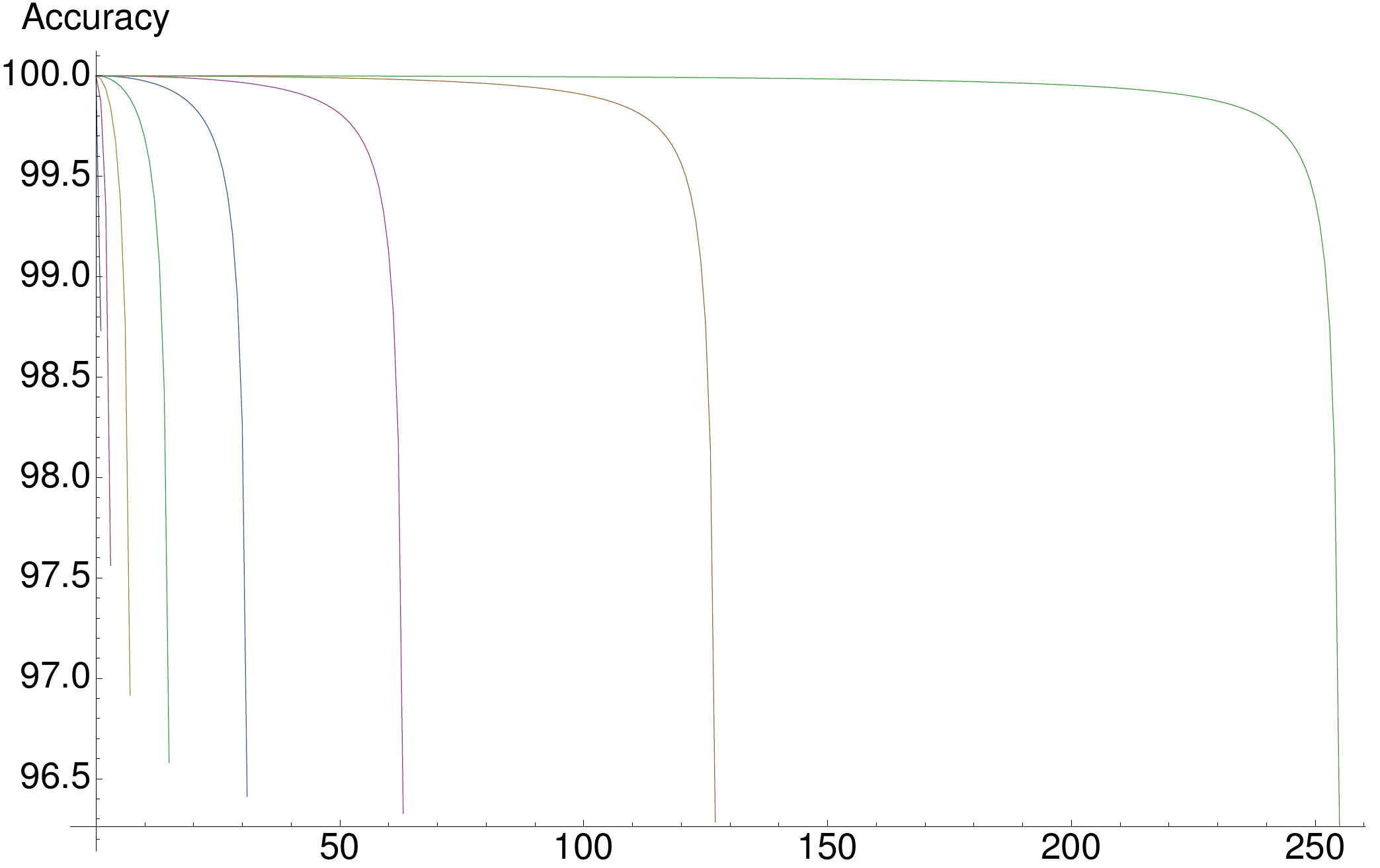}};
        \node[rotate=90, anchor=west, right] at (-0.5,2) {\small Accuracy} ;
        \node[anchor=west, right] at (7.9,0.1) {$s$} ;
\end{tikzpicture}
      \vspace{-0.7cm}
    \caption{Numerical accuracy (\ie\ 100 minus the absolute percentage difference between ${\rm FWHM}^{s > 0}_{\theta}$ and ${\rm FWHM}^{s=0}_{\theta}$) against the spin value $s$. Different curves (from left to right) corresponds to different fixed $\el= 2^{p}$ cases, where integer $p$ runs from 1 to 8. The results shows that ${\rm FWHM}^{s > 0}_{\theta}$ (\ie\ the spin setting) remains very close to ${\rm FWHM}^{s=0}_{\theta}$ (\ie\ the scalar setting, for which typical curvelet parabolic scaling relation has been shown to hold). This suggests that the parabolic scaling relation should hold for at least $s \simeq  \floor{\el/2}$, conservatively speaking. For $s$ approaching $\el$, the parabolic scaling relation still holds to good approximation. 
Even in the worst case when $s=\el-1$, the error may be tolerable (\eg\ the error is within $5\%$ at $\el = 256$).}
    \label{fig:spin_setting_FWHM}
\end{figure}

As noted in \sectn{\ref{subsec:curveletConstruct}}, 
due to the asymmetric property of $d^\el_{\el (-s)}(\theta)$ about $\theta_{\rm max}$ as $s \rightarrow (\el -1)$, 
the parabolic scaling relation which has been shown to hold for cases $s=0$ and $s \ll \el$ may start to break down. 
To investigate at which $s$ the offset from parabolic scaling becomes important, we evaluate the absolute percentage difference between ${\rm FWHM}^{s > 0}_{\theta}$ and ${\rm FWHM}^{s=0}_{\theta}$ at a set of test values $\el = 2^{p}$ where $p$ runs from 1 to 8. 
Our empirical results are plotted in \fig{\ref{fig:spin_setting_FWHM}}, from which one can see that for up to $s \simeq \el/2$ in all test-$\el$ cases, 
${\rm FWHM}^{s > 0}_{\theta}$ (\ie\ the spin setting) remains very close to ${\rm FWHM}^{s=0}_{\theta}$, 
with percentage error $<0.05\%$. Hence, $s \simeq \floor{\el/2}$ can serve as a very conservative limit 
for which the typical curvelet parabolic scaling relation remains to hold. 
Nevertheless, even for $s$ approaching $\el$, the parabolic scaling relation holds to good approximation. 

\ifCLASSOPTIONcaptionsoff
  \newpage
\fi



%

\bibliographystyle{IEEEtran}
\bibliography{bib_myname,bib_journal_names_long,references}

\begin{thebibliography}{10}
\providecommand{\url}[1]{#1}
\csname url@rmstyle\endcsname
\providecommand{\newblock}{\relax}
\providecommand{\bibinfo}[2]{#2}
\providecommand\BIBentrySTDinterwordspacing{\spaceskip=0pt\relax}
\providecommand\BIBentryALTinterwordstretchfactor{4}
\providecommand\BIBentryALTinterwordspacing{\spaceskip=\fontdimen2\font plus
\BIBentryALTinterwordstretchfactor\fontdimen3\font minus
  \fontdimen4\font\relax}
\providecommand\BIBforeignlanguage[2]{{%
\expandafter\ifx\csname l@#1\endcsname\relax
\typeout{** WARNING: IEEEtran.bst: No hyphenation pattern has been}%
\typeout{** loaded for the language `#1'. Using the pattern for}%
\typeout{** the default language instead.}%
\else
\language=\csname l@#1\endcsname
\fi
#2}}

\bibitem{antoine:1999}
J.-P. Antoine and P.~Vandergheynst, ``Wavelets on the 2-sphere: a group
  theoretical approach,'' \emph{Applied Comput.\ Harm.\ Anal.}, vol.~7, pp.
  1--30, 1999.

\bibitem{antoine:1998}
------, ``Wavelets on the n-sphere and related manifolds,'' \emph{J.\ Math.\
  Phys.}, vol.~39, no.~8, pp. 3987--4008, 1998.

\bibitem{mcewen:szip}
J.~D. McEwen, Y.~Wiaux, and D.~M. Eyers, ``Data compression on the sphere,''
  \emph{Astron.\ \& Astrophys.}, vol. 531, no. A98, 2011.

\bibitem{narcowich:2006}
F.~J. Narcowich, P.~Petrushev, and J.~D. Ward, ``Localized tight frames on
  spheres,'' \emph{SIAM J. Math. Anal.}, vol.~38, no.~2, pp. 574--594, 2006.

\bibitem{baldi:2009}
P.~{Baldi}, G.~{Kerkyacharian}, D.~{Marinucci}, and D.~{Picard}, ``{Asymptotics
  for spherical needlets},'' \emph{Ann.\ Stat.}, vol. 37 No.3, pp. 1150--1171,
  2009.

\bibitem{marinucci:2008}
D.~{Marinucci}, D.~{Pietrobon}, A.~{Balbi}, P.~{Baldi}, P.~{Cabella},
  G.~{Kerkyacharian}, P.~{Natoli}, D.~{Picard}, and N.~{Vittorio}, ``{Spherical
  needlets for cosmic microwave background data analysis},'' \emph{Mon.\ Not.\
  Roy.\ Astron.\ Soc.}, vol. 383, pp. 539--545, 2008.

\bibitem{sanz:2006}
J.~L. Sanz, D.~Herranz, M.~L{\'o}pez-Caniego, and F.~Arg{\"u}eso, ``Wavelets on
  the sphere -- application to the detection problem,'' in \emph{Proc. 14th
  Eur. Signal Process. Conf.}, Sept. 2006, pp. 1--5.

\bibitem{wiaux:2005}
Y.~Wiaux, L.~Jacques, and P.~Vandergheynst, ``Correspondence principle between
  spherical and {E}uclidean wavelets,'' \emph{Astrophys.\ J.}, vol. 632, pp.
  15--28, 2005.

\bibitem{mcewen:2006:cswt2}
\BIBentryALTinterwordspacing
J.~D. {McEwen}, M.~P. {Hobson}, and A.~N. {Lasenby}, ``{A directional
  continuous wavelet transform on the sphere},'' \emph{ArXiv e-prints}, Sept.
  2006. [Online]. Available:
  \url{http://adsabs.harvard.edu/abs/2006astro.ph..9159M}
\BIBentrySTDinterwordspacing

\bibitem{wiaux:2007:sdw}
Y.~Wiaux, J.~D. McEwen, P.~Vandergheynst, and O.~Blanc, ``Exact reconstruction
  with directional wavelets on the sphere,'' \emph{Mon.\ Not.\ Roy.\ Astron.\
  Soc.}, vol. 388, no.~2, pp. 770--788, 2008.

\bibitem{leistedt:s2let_axisym}
B.~Leistedt, J.~D. McEwen, P.~Vandergheynst, and Y.~Wiaux, ``{S2LET}: A code to
  perform fast wavelet analysis on the sphere,'' \emph{Astron.\ \& Astrophys.},
  vol. 558, no. A128, pp. 1--9, 2013.

\bibitem{mcewen:2013:waveletsxv}
J.~D. McEwen, P.~Vandergheynst, and Y.~Wiaux, ``On the computation of
  directional scale-discretized wavelet transforms on the sphere,'' in
  \emph{SPIE Wavelets and Sparsity XV}, vol. 8858, 2013, pp. 88\,580I--1--13.

\bibitem{mcewen:s2let_localisation}
\BIBentryALTinterwordspacing
J.~D. McEwen, C.~Durastanti, and Y.~Wiaux, ``Localisation of directional
  scale-discretised wavelets on the sphere,'' \emph{Applied and Computational
  Harmonic Analysis}, 2016. [Online]. Available:
  \url{http://www.sciencedirect.com/science/article/pii/S1063520316000324}
\BIBentrySTDinterwordspacing

\bibitem{mcewen:s2let_spin}
\BIBentryALTinterwordspacing
J.~D. McEwen, B.~Leistedt, M.~B{\"u}ttner, H.~V. Peiris, and Y.~Wiaux,
  ``Directional spin wavelets on the sphere,'' \emph{ArXiv e-prints}, Sept.
  2015. [Online]. Available:
  \url{http://adsabs.harvard.edu/abs/2015arXiv150906749M}
\BIBentrySTDinterwordspacing

\bibitem{mcewen:s2let_ridgelets}
\BIBentryALTinterwordspacing
J.~D. {McEwen}, ``{Ridgelet transform on the sphere},'' \emph{ArXiv e-prints},
  Oct. 2015. [Online]. Available:
  \url{http://adsabs.harvard.edu/abs/2015arXiv151001595M}
\BIBentrySTDinterwordspacing

\bibitem{mcewen:2008:fsi}
J.~D. McEwen and A.~M.~M. Scaife, ``Simulating full-sky interferometric
  observations,'' \emph{Mon.\ Not.\ Roy.\ Astron.\ Soc.}, vol. 389, no.~3, pp.
  1163--1178, 2008.

\bibitem{michailovich:2010a}
O.~Michailovich and Y.~Rathi, ``On approximation of orientation distributions
  by means of spherical ridgelets,'' \emph{IEEE Trans.\ Image Proc.}, vol.~19,
  no.~2, pp. 461--477, Feb 2010.

\bibitem{starck:2006}
J.-L. {Starck}, Y.~{Moudden}, P.~{Abrial}, and M.~{Nguyen}, ``{Wavelets,
  ridgelets and curvelets on the sphere},'' \emph{Astron.\ \& Astrophys.}, vol.
  446, pp. 1191--1204, Feb. 2006.

\bibitem{geller:2008}
D.~{Geller}, F.~K. {Hansen}, D.~{Marinucci}, G.~{Kerkyacharian}, and
  D.~{Picard}, ``{Spin needlets for cosmic microwave background polarization
  data analysis},'' \emph{Phys.\ Rev.\ D.}, vol.~78, p. 123533, Dec 2008.

\bibitem{geller:2010:sw}
D.~{Geller} and D.~{Marinucci}, ``{Spin Wavelets on the Sphere},'' \emph{J.\
  Fourier Anal.\ and Appl.}, vol.~16, no.~6, pp. 840–--884, Nov. 2010.

\bibitem{geller:2010}
------, ``{Mixed Needlets},'' \emph{J. Math. Anal. Appl.}, vol. 375, no.~2, pp.
  610–--630, June 2011.

\bibitem{vielva:2004}
P.~Vielva, E.~Mart\'{\i}nez-Gonz\'{a}lez, R.~B. Barreiro, J.~L. Sanz, and
  L.~Cay\'{o}n, ``Detection of non-{G}aussianity in the {WMAP} 1-year data
  using spherical wavelets,'' \emph{Astrophys.\ J.}, vol. 609, pp. 22--34,
  2004.

\bibitem{mcewen:2008:ng}
J.~D. McEwen, M.~P. Hobson, A.~N. Lasenby, and D.~J. Mortlock, ``A
  high-significance detection of non-{G}aussianity in the {WMAP} 5-year data
  using directional spherical wavelets,'' \emph{Mon.\ Not.\ Roy.\ Astron.\
  Soc.}, vol. 388, no.~2, pp. 659--662, 2008.

\bibitem{mcewen:2006:bianchi}
------, ``Non-{G}aussianity detections in the {B}ianchi {VII{$_h$}} corrected
  {WMAP} 1-year data made with directional spherical wavelets,'' \emph{Mon.\
  Not.\ Roy.\ Astron.\ Soc.}, vol. 369, pp. 1858--1868, 2006.

\bibitem{delabrouille:2009}
J.~{Delabrouille}, J.-F. {Cardoso}, M.~{Le Jeune}, M.~{Betoule}, G.~{Fay}, and
  F.~{Guilloux}, ``{A full sky, low foreground, high resolution CMB map from
  WMAP},'' \emph{Astron.\ \& Astrophys.}, vol. 493, pp. 835--857, Jan. 2009.

\bibitem{vielva:2005}
P.~{Vielva}, E.~{Mart{\'{\i}}nez-Gonz{\'a}lez}, and M.~{Tucci},
  ``{Cross-correlation of the cosmic microwave background and radio galaxies in
  real, harmonic and wavelet spaces: detection of the integrated Sachs-Wolfe
  effect and dark energy constraints},'' \emph{Mon.\ Not.\ Roy.\ Astron.\
  Soc.}, vol. 365, pp. 891--901, 2006.

\bibitem{vielva:2006}
P.~{Vielva}, Y.~{Wiaux}, E.~{Mart{\'\i}nez-Gonz\'alez}, and P.~{Vandergheynst},
  ``{Steerable wavelet analysis of CMB structures alignment},'' \emph{New
  Astronomy Review}, vol.~50, pp. 880--888, 2006.

\bibitem{lan:2008}
X.~{Lan} and D.~{Marinucci}, ``{The needlets bispectrum},'' \emph{Electronic
  Journal of Statistics}, vol.~2, pp. 332--367, 2008.

\bibitem{mcewen:2007:isw2}
J.~D. McEwen, Y.~Wiaux, M.~P. Hobson, P.~Vandergheynst, and A.~N. Lasenby,
  ``Probing dark energy with steerable wavelets through correlation of {WMAP}
  and {NVSS} local morphological measures,'' \emph{Mon.\ Not.\ Roy.\ Astron.\
  Soc.}, vol. 384, no.~4, pp. 1289--1300, 2008.

\bibitem{pietrobon:2006}
D.~Pietrobon, A.~Balbi, and D.~Marinucci, ``Integrated {S}achs-{W}olfe effect
  from the cross-correlation of {WMAP} 3-year and {NVSS}: new results and
  constraints on dark energy,'' \emph{Phys.\ Rev.\ D.}, vol.~74, no.~4, p.
  043524, Aug. 2006.

\bibitem{planck2013-p06}
{Planck Collaboration XII}, ``{\textit{Planck} 2013 results. XII. Diffuse
  component separation},'' \emph{Astron.\ \& Astrophys.}, vol. 571, no. A12,
  Nov. 2014.

\bibitem{planck2013-p09}
{Planck Collaboration XXIII}, ``{\textit{Planck} 2013 results. XXIII. Isotropy
  and statistics of the CMB},'' \emph{Astron.\ \& Astrophys.}, vol. 571, no.
  A23, Nov. 2014.

\bibitem{planck2013-p20}
{Planck Collaboration XXV}, ``{\textit{Planck} 2013 results. XXV. Searches for
  cosmic strings and other topological defects},'' \emph{Astron.\ \&
  Astrophys.}, vol. 571, no. A25, Nov. 2014.

\bibitem{planck2015-p4}
\BIBentryALTinterwordspacing
{Planck Collaboration IX}, ``{\textit{Planck} 2015 results. IX. Diffuse
  component separation: CMB maps},'' \emph{ArXiv e-prints}, Feb. 2015.
  [Online]. Available: \url{http://adsabs.harvard.edu/abs/2015arXiv150205956P}
\BIBentrySTDinterwordspacing

\bibitem{bobin:2013}
J.~{Bobin}, J.-L. {Starck}, F.~{Sureau}, and S.~{Basak}, ``{Sparse component
  separation for accurate cosmic microwave background estimation},''
  \emph{Astron.\ \& Astrophys.}, vol. 550, no. A73, Feb. 2013.

\bibitem{mcewen:2006:review}
J.~D. McEwen, P.~Vielva, Y.~Wiaux., R.~B. Barreiro, L.~Cay\'on, M.~P. Hobson,
  A.~N. Lasenby, E.~Mart{\'\i}nez-Gonz\'alez, and J.~L. Sanz, ``Cosmological
  applications of a wavelet analysis on the sphere,'' \emph{J.\ Fourier Anal.\
  and Appl.}, vol.~13, no.~4, pp. 495--510, 2007.

\bibitem{schmitt:2012}
J.~{Schmitt}, J.~L. {Starck}, J.~M. {Casandjian}, J.~{Fadili}, and
  I.~{Grenier}, ``{Multichannel Poisson denoising and deconvolution on the
  sphere: application to the Fermi Gamma-ray Space Telescope},'' \emph{Astron.\
  \& Astrophys.}, vol. 546, no. A114, Oct. 2012.

\bibitem{Holschneider:2003}
M.~Holschneider, A.~Chambodut, and M.~Mandea, ``From global to regional
  analysis of the magnetic field on the sphere using wavelet frames,''
  \emph{Physics of the Earth and Planetary Interiors}, vol. 135, no. 2--3, pp.
  107--124, 2003, magnetic Field Modelling.

\bibitem{audet:2014}
P.~Audet, ``Toward mapping the effective elastic thickness of planetary
  lithospheres from a spherical wavelet analysis of gravity and topography,''
  \emph{Phys.\ Earth Planet In.}, vol. 226, no.~0, pp. 48--82, 2014.

\bibitem{schmidt:2006}
M.~Schmidt, S.-C. Han, J.~Kusche, L.~Sanchez, and C.~K. Shum, ``Regional
  high-resolution spatiotemporal gravity modeling from grace data using
  spherical wavelets,'' \emph{Geophysical Research Letters}, vol.~33, no.~8,
  2006, art. no. L08403.

\bibitem{simons:2011}
F.~J. {Simons}, I.~{Loris}, G.~{Nolet}, I.~C. {Daubechies}, S.~{Voronin}, J.~S.
  {Judd}, P.~A. {Vetter}, J.~{Charl{\'e}ty}, and C.~{Vonesch}, ``{Solving or
  resolving global tomographic models with spherical wavelets, and the scale
  and sparsity of seismic heterogeneity},'' \emph{Geophysical Journal
  International}, vol. 187, pp. 969--988, Nov. 2011.

\bibitem{Rathi:2011tq}
Y.~Rathi, O.~Michailovich, K.~Setsompop, S.~Bouix, M.~E. Shenton, and C.-F.
  Westin, ``{Sparse multi-shell diffusion imaging},'' \emph{Medical image
  computing and computer-assisted intervention : MICCAI. International
  Conference on Medical Image Computing and Computer-Assisted Intervention},
  vol.~14, no. 0 2, pp. 58--65, 2011.

\bibitem{candes:1999:ridgelets}
E.~J. Cand{\`e}s and D.~L. Donoho, ``Ridgelets: a key to higher-dimensional
  intermittency?'' \emph{Philosophical Transactions of the Royal Society of
  London A: Mathematical, Physical and Engineering Sciences}, vol. 357, no.
  1760, pp. 2495--2509, 1999.

\bibitem{candes:1999curvelets}
E.~J. Candes, D.~L. Donoho, \emph{et~al.}, \emph{Curvelets: A surprisingly
  effective nonadaptive representation for objects with edges}.\hskip 1em plus
  0.5em minus 0.4em\relax Standford, CA, USA: Dept. of Statistics, Stanford
  University, 1999.

\bibitem{Candes:2004}
E.~J. Cand{\`e}s and D.~L. Donoho, ``New tight frames of curvelets and optimal
  representations of objects with piecewise c2 singularities,''
  \emph{Communications on Pure and Applied Mathematics}, vol.~57, no.~2, pp.
  219--266, 2004.

\bibitem{Candes:2005bg}
------, ``Continuous curvelet transform: I. resolution of the wavefront set,''
  \emph{Applied and Computational Harmonic Analysis}, vol.~19, no.~2, pp.
  162--197, Sept. 2005.

\bibitem{Candes:2005bg2}
------, ``Continuous curvelet transform: Ii. discretization and frames,''
  \emph{Applied and Computational Harmonic Analysis}, vol.~19, no.~2, pp. 198
  -- 222, 2005.

\bibitem{Ma:2009tj}
J.~Ma and G.~Plonka, ``{A review of curvelets and recent applications},''
  \emph{IEEE Signal Processing Magazine}, pp. 118€"--133, 2010.

\bibitem{Fadili:2012ht}
J.~Fadili and J.-L. Starck, ``{Curvelets and Ridgelets},'' in
  \emph{Computational Complexity}.\hskip 1em plus 0.5em minus 0.4em\relax New
  York, NY: Springer New York, Jan. 2012, pp. 754 --773.

\bibitem{Starck:2005hl}
J.-L. {Starck}, Y.~{Moudden}, P.~{Abrial}, and M.~{Nguyen}, ``{Wavelets,
  ridgelets and curvelets on the sphere},'' \emph{Astron.\ \& Astrophys.}, vol.
  446, pp. 1191--1204, Feb. 2006.

\bibitem{gorski:2005}
K.~M. G\'{o}rski, E.~Hivon, A.~J. Banday, B.~D. Wandelt, F.~K. Hansen,
  M.~Reinecke, and M.~Bartelmann, ``Healpix -- a framework for high resolution
  discretization and fast analysis of data distributed on the sphere,''
  \emph{Astrophys.\ J.}, vol. 622, pp. 759--771, 2005.

\bibitem{driscoll:1994}
J.~R. Driscoll and D.~M.~J. Healy, ``Computing {F}ourier transforms and
  convolutions on the sphere,'' \emph{Adv.\ Appl.\ Math.}, vol.~15, pp.
  202--250, 1994.

\bibitem{mcewen:fssht}
J.~D. McEwen and Y.~Wiaux, ``A novel sampling theorem on the sphere,''
  \emph{IEEE Trans.\ Sig.\ Proc.}, vol.~59, no.~12, pp. 5876--5887, 2011.

\bibitem{mcewen:so3}
J.~D. {McEwen}, M.~{Buttner}, B.~{Leistedt}, H.~V. {Peiris}, and Y.~{Wiaux},
  ``{A Novel Sampling Theorem on the Rotation Group},'' \emph{IEEE Signal
  Processing Letters}, vol.~22, pp. 2425--2429, Dec. 2015.

\bibitem{varshalovich:1989}
D.~A. Varshalovich, A.~N. Moskalev, and V.~K. Khersonskii, \emph{Quantum theory
  of angular momentum}.\hskip 1em plus 0.5em minus 0.4em\relax Singapore: World
  Scientific, 1989.

\bibitem{goldberg:1967}
J.~N. Goldberg, A.~J. Macfarlane, E.~T. Newman, F.~Rohrlich, and E.~C.~G.
  Sudarshan, ``Spin-$s$ spherical harmonics and $\eth$,'' \emph{J.\ Math.\
  Phys.}, vol.~8, no.~11, pp. 2155--2161, 1967.

\bibitem{FFTW05}
M.~Frigo and S.~G. Johnson, ``The design and implementation of {FFTW3},''
  \emph{Proceedings of the IEEE}, vol.~93, no.~2, pp. 216--231, 2005, special
  issue on ``Program Generation, Optimization, and Platform Adaptation''.

\bibitem{PaulDebevec:lightprobe}
P.~Debevec, ``Rendering synthetic objects into real scenes: Bridging
  traditional and image-based graphics with global illumination and high
  dynamic range photography,'' in \emph{Proc. 25th Annu. Conf. Comput. Graph.
  Interactive Techn.}\hskip 1em plus 0.5em minus 0.4em\relax Orlando, Florida,
  July 1998, pp. 189--198.

\bibitem{leistedt:flaglets}
B.~Leistedt and J.~D. McEwen, ``Exact wavelets on the ball,'' \emph{IEEE
  Trans.\ Sig.\ Proc.}, vol.~60, no.~12, pp. 6257--6269, 2012.

\bibitem{mcewen:flaglets_sampta}
J.~D. McEwen and B.~Leistedt, ``{F}ourier-{L}aguerre transform, convolution and
  wavelets on the ball,'' in \emph{Proc. 10th Int. Conf. Sampling Theory and
  Appl.}, 2013, pp. 329--333.

\end{thebibliography}




%








\end{document}